\documentstyle[12pt]{article}
\def\boxit#1{
{\vbox{\hrule\hbox{\vrule\kern3pt
\vbox{\kern3pt{\vspace{1.8ex}#1}\kern3pt}\kern3pt\vrule}\hrule}}}
\newfont{\titlesl}{cmsl10 scaled\magstep4}
\def\Title#1#2{
\hfill\hbox{#1}
\vskip 3pt
\hfill\hbox{hep-th/9509022}
\begin{center}
\baselineskip=2\baselineskip
\titlesl #2
\end{center}
\baselineskip=.5\baselineskip
\normalsize\vskip .5in}
\def\authorinfo#1#2#3{\centerline{\large\sc#1\footnote{e-mail:
#2}\vspace{12pt} } \centerline{\normalsize\em#3}\vspace{12pt}}
\def\hcentering{%
	\oddsidemargin=211mm
	\advance\oddsidemargin-\textwidth
	\oddsidemargin=0.5\oddsidemargin
	\advance\oddsidemargin-1in
	\evensidemargin=\oddsidemargin}
\def\vcentering{%
	\topmargin=297mm
	\advance\topmargin-\textheight
	\topmargin=0.5\topmargin
	\headheight=0.4\topmargin
	\topmargin=0.2\topmargin
	\advance\topmargin-1in
	\headsep=\headheight
	\footskip=\headsep
	\footheight=\headheight}
\makeatletter
\def\section{\@startsection {section}{1}{\z@}{-3.5ex plus -1ex minus
 -.2ex}{2.3ex plus .2ex}{\large\bf}}
\def\subsection{\@startsection{subsection}{2}{\z@}{-3.25ex plus -1ex minus
 -.2ex}{1.5ex plus .2ex}{\normalsize\bf}}
\makeatother

\makeatletter
\@addtoreset{equation}{section}

\makeatother

\newcommand{\nc}{\newcommand}
\newcommand{\rnc}{\renewcommand}
\def\itembox#1{\hbox to 3em{\hfill(#1)\hfill}}
\def\itembra#1{\hbox to 3em{\hfill[#1]\hfill}}

\nc{\be}{\begin{equation}}
\nc{\ee}{\end{equation}}
\nc{\bea}{\begin{eqnarray}}
\nc{\eea}{\end{eqnarray}}
\def\eqref#1{(\ref{#1})}

\newtheorem{mytheorem}{Theorem}[section]
\newtheorem{mylemma}[mytheorem]{Lemma}
\newtheorem{mycorol}[mytheorem]{Corollaly}

\rnc{\a}{\alpha}
\rnc{\b}{\beta}
\rnc{\gg}{\gamma}
\rnc{\d}{\delta}
\nc{\e}{\eta}
\nc{\eb}{\bar{\eta}}
\nc{\ep}{\epsilon}
\nc{\f}{\phi}
\nc{\fb}{\bar{\phi}}
\nc{\vf}{\varphi}
\nc{\p}{\psi}
\nc{\bpsi}{\bar{\psi}}
\rnc{\c}{\chi}
\nc{\cb}{\bar{\c}}
\nc{\la}{\lambda}
\nc{\m}{\mu}
\nc{\n}{\nu}
\nc{\bmu}{\bar{\mu}}
\rnc{\o}{\omega}
\nc{\Om}{\Omega}
\rnc{\t}{\theta}
\nc{\eps}{\epsilon}
\nc{\veps}{\varepsilon}
\nc{\bxi}{\bar{\xi}}
\def\Dt{\Delta t}
\makeatletter
\def\mathhexbox{\protect\mathhexbox@}
\def\mathhexbox@#1#2#3{\relax
\ifmmode\mathpalette{}{\m@th\mathchar"#1#2#3}%
\else\leavevmode\hbox{$\m@th\mathchar"#1#2#3$}\fi}
\def\text#1{%
\relax
\ifmmode %
\mathchoice
{\hbox{\everymath{\displaystyle}\rm #1}}%
{\hbox{\everymath{\textstyle}\rm #1}}%
{\hbox{\everymath{\scriptstyle}%
\def\prm{\fam\z@ \the\scriptfont\z@ \relax}%
\def\pit{\fam\itfam \the\scriptfont\itfam \relax}%
\rm #1}%
}%
{\hbox{\everymath{\scriptscriptstyle}%
\def\prm{\fam\z@ \the\scriptscriptfont\z@ \relax}%
\def\pit{\fam\itfam \the\scriptscriptfont\itfam \relax}%
\rm #1}%
}%
\else %
\leavevmode\hbox{#1}%
\fi
}
\def\bbox#1{%
\leavevmode\text{%
\textfont0 \the\textfont\bffam
\scriptfont0 \the\scriptfont\bffam
\scriptscriptfont0 \the\scriptscriptfont\bffam
\@temptokena\everymath \boldmath \everymath\@temptokena
{$\m@th\relax#1$}%
}%
}
\font\fivbf=cmbx5 \font\sixbf=cmbx6 \font\sevbf=cmbx7 \font\egtbf=cmbx8
\expandafter\def\expandafter\ixpt\expandafter{\ixpt
\scriptfont\bffam\sixbf \scriptscriptfont\bffam\fivbf}
\expandafter\def\expandafter\xpt\expandafter{\xpt
\scriptfont\bffam\sevbf \scriptscriptfont\bffam\fivbf}
\expandafter\def\expandafter\xipt\expandafter{\xipt
\scriptfont\bffam\egtbf \scriptscriptfont\bffam\sixbf}
\expandafter\def\expandafter\xiipt\expandafter{\xiipt
\scriptfont\bffam\egtbf \scriptscriptfont\bffam\sixbf}
\expandafter\def\expandafter\xivpt\expandafter{\xivpt
\scriptfont\bffam\tenbf \scriptscriptfont\bffam\sevbf}
\makeatother

\rnc{\S}{\Sigma}
\nc{\Sg}{\Sigma_{g}}
\nc{\Sa}{\S\times\{0\}}
\nc{\Sb}{\S\times\{1\}}
\nc{\SI}{\S\times I}
\nc{\SS}{\S\times S^{1}}
\nc{\M}{{\cal M}}
\nc{\G}{{\cal G}}
\let\Cal=\cal
\nc{\trac}[2]{{\textstyle\frac{#1}{#2}}}

\nc{\ex}[1]{{\rm e}^{\,\textstyle#1}}

\nc{\mat}[4]{\left(\begin{array}{cc}#1&#2\\#3&#4\end{array}\right)}

\nc{\mtc}[9]{\left(\begin{array}{ccc}#1&#2&#3\\#4&#5&#6\\#7&#8&#9%
\end{array}\right)}

\def\tr{\mathop{\rm tr}\nolimits}
\def\Tr{\mathop{\rm Tr}\nolimits}
\def\Un#1{\mathop{\rm U}(#1)}
\def\SUn#1{\mathop{\rm SU}(#1)}
\def\Mn#1{M(#1;\bbox{C})}
\def\Mnm#1#2{M(#1,#2;\bbox{C})}
\def\Hn#1{\mathop{\rm H}(#1)}
\let\Bbb=\bbox
\nc{\ra}{\rightarrow}
\nc{\Ra}{\Rightarrow}
\nc{\Lra}{\Leftrightarrow}
\nc{\ot}{\otimes}
\rnc{\ss}{\subset}
\nc{\nul}{\noindent\underline}
\nc{\ad}{\mbox{ad}}
\nc{\non}{\nonumber\\}
\newfont{\tenmb}{cmmib10}
\newfont{\elemb}{cmmib10 scaled\magstephalf}
\newfont{\twemb}{cmmib10 scaled\magstep1}
\nc{\mib}[1]{\mbox{\boldmath $#1$}}
\def\boldkey#1{\mib{#1}}

\def\sgn#1{\mbox{\rm sgn}(#1)}
\def\grsmn#1#2{G_{#1,#2}}
\def\nxi{\displaystyle 1_{n}+\xi^{\dagger}\xi}
\def\mxi{\displaystyle 1_{N-n}+\xi\xi^{\dagger}}
\def\dfrac#1#2{{{\displaystyle #1}\over{\displaystyle #2}}}
\def\dchoose#1#2{\displaystyle#1\choose\displaystyle#2}
\def\xidag{\xi^{\dagger}}
\def\delxi#1#2{{{\partial #1}\over{\partial \xi_{#2}}}}
\def\delbxi#1#2{{{\partial #1}\over{\partial \bxi_{#2}}}}
\begin{document}
\hcentering\vcentering
\begin{titlepage}
%%%
\Title{KYUSHU-HET-27}
{Coherent states over Grassmann manifolds\\
and\\
the WKB-exactness in path integral
}
%%%
\baselineskip=1.25\baselineskip
%%%
\authorinfo{Kazuyuki FUJII}{fujii@yokohama-cu.ac.jp}
{Department of Mathematics, Yokohama City
University, Yokohama 236, Japan,}
\authorinfo{Taro KASHIWA}{taro1scp@mbox.nc.kyushu-u.ac.jp}
{Department of Physics, Kyushu
University, Fukuoka 812-81, Japan,}
\authorinfo{Seiji SAKODA}{a00501@cc.hc.keio.ac.jp}
{Department of Physics, Hiyoshi,
Keio  University, Yokohama 223, Japan}
%%%
\setcounter{footnote}{0}

\baselineskip=.8\baselineskip

%%%
\begin{abstract}
\(\Un{N}\) coherent states over Grassmann manifold,
\(\grsmn{N}{n}\simeq\Un{N}/ (\Un{n}\times \Un{N-n})\), are formulated
to be able to argue the WKB-exactness, so called the localization of
Duistermaat-Heckman, in the path integral representation of a character
formula.
The exponent in the path integral formula is proportional to an integer
\(k\) that labels the \(\Un{N}\) representation. Thus when
\(k \rightarrow\infty\) a usual semiclassical approximation, by regarding
\(k \sim 1 / \hbar\), can be performed yielding to a desired conclusion. The
mechanism of the localization is uncovered by means of a view from an extended
space realized by the Schwinger boson technique.
\end{abstract}
\end{titlepage}
%%%
\section{Introduction}
\label{intr}

In any physical situation it is often difficult to find an exact
response, therefore, some approximation method must be
employed. Apart from the well-known perturbation theories, the
Wentzel-Kramers-Brillouin(WKB) approximation, known as the
\(\hbar\)-expansion,
seems most suitable to the path integral formalism; since the exponent in
the path integral
representation is usually given by a quantity divided by Planck's
constant \(\hbar\). These
approximation methods can be straightforwardly performed without
specifying a path
measure rigorously, which apparently has been a main reason that path
integral plays a
major role in modern physics.  On the other hand, due to this facileness
there always
accompanies some skepticism, such as the problem of operator
ordering\cite{KSZ} since
only \(c\)-numbers appear, or a vague relationship of change of variables to
the canonical
operator formalism\cite{FK}. What we have learned through various efforts
is that path
integral can produce reliable as well as  consistent results under the
time slicing method. A
simple example can be seen as follows: take a bosonic oscillator, defined
by a Hamiltonian
\(H\equiv\omega a^{\dagger}a\) with the algebra
\(
[a,a^{\dagger}]={\bf I},\
[{\bf I},a]=[{\bf I},a^{\dagger}]=0\ ,
\)
and calculate the quantity \(\text{Tr} e^{-i H T }\). First write
the exponential operator such that
\be
\text{Tr} e^{-iHT} = \lim_{M \rightarrow \infty} \text{Tr}\left( {\bf I}
-  i \Delta t H
\right)^M ; \quad  \Delta t \equiv {T \over M} \quad, \label{timeslice}
\ee
which is the starting point of the time slicing method. With the aid of
canonical coherent
states\cite{KS}, \eqref{timeslice} becomes
\bea
 \!\!\!\! \text{Tr} e^{-i\omega a^{\dagger}a T }
& = &
\lim_{M\rightarrow \infty}
\int\limits_{\text{PBC}}\! \prod_{j=1}^{M}{ dz(j)d\bar{z}(j) \over \pi}
\nonumber \\
& \times  & \exp\left[
- \sum_{k=1}^{M} \left\{{\bar z}(k) (z(k)-z(k-1)) +  i\omega
\Delta t{\bar z}(k)
z(k-1)
\right\}\right]  ,
\label{discreterep}
\eea
where ``PBC" denotes
\(z(0) = z(M)\) and \(dzd\bar{z}\equiv d\text{Re}(z)d\text{Im}(z)\).
By taking a formal limit, \(M\rightarrow \infty\),
\eqref{discreterep} yields to the continuum representation,
\be
\text{Tr}e^{-i\omega a^{\dagger}a T }
 \rightarrow  \int\limits_{\text{PBC}}\!
\prod_{0\le t\le T}dzd\bar{z}
\exp\left\{
- \int_0^T  dt \left(   {\bar z} \dot z + i\omega {\bar z}z
\right)  \right\}  \quad, \label{continrep}
\ee
where ``PBC'' reads \(z(T) = z(0)\) in this case. In spite of the formal
limit, we still
could endow a meaning with the functional measure by means of the
functional
determinant:
\be
\eqref{continrep} \equiv {\det \left( {d \over dt} + i
\omega \right)}^{-1} =  {1\over{2i\sin(\omega T/2)}}  \quad,
\label{intrf}
\ee
with the aid of the \(\zeta\)-function regularization. The result does not,
however, match to the correct one given by \eqref{discreterep}
\be
{1\over{2i\sin(\omega T/2)}} \neq
{1\over{1-e^{-i\omega T}}}  = {e^{i
\omega T/2} \over {2i\sin(\omega T/2)} } \quad.
\label{intrh}
\ee
Therefore we must pay the price whenever we have adopted the continuum
path integral
representation which would apparently be suitable for a geometrical
treatment.

In some situation an approximation scheme happens to lead to an exact
answer.  The harmonic oscillator is WKB-exact, because of the integration
being Gaussian. (The cross section of the Coulomb interaction is another
well-known example\cite{Messiah}, which furthermore reveals that
the Born approximation yields the exact result.) In recent  years, however,
a new possibility of finding the WKB exactness has been opened
up\cite{SN,AFS,VNS,RR,FFKS1,FFKS2} being inspired by the
Duistermaat-Heckman's (D-H) theorem\cite{DH,PKN,MFA}. A key word to
understand these new classes of the WKB exactness would be `{\em
localization\/}' \cite{BT,PKN2,BVN}, commonly understood in terms of
equivariant cohomology\cite{KN}.

Inspired by these facts, we have established the WKB
exactness of path integral obtained through the generalized coherent
states\cite{Perel} in
cases of \(\bbox{C}P^{1}\) \cite{FFKS1} and \(\bbox{C}P^{N}\)\cite{FFKS2}
as well
as their  noncompact counter  parts in the
foregoing papers.  Even if other representation is
adopted\cite{TK} for the
\(\bbox{C}P^{1}\) case to give the Nielsen-Rohrlich form\cite{NR}, the
same localization
has been clarified\cite{FKNS}. As a natural generalization in this paper,
we try to understand
the same phenomena in the case of the Grassmann manifold,
\(\grsmn{N}{n}\simeq\Un{N}/(\Un{n}\times\Un{N-n})\). To this end, we
need to build up coherent states of \(\Un{N}\) over \(\grsmn{N}{n}\).

The plan of the paper is as follows. An interpretation of the D-H theorem,
stated in terms of
finite dimensional integrations, is presented in section \ref{D-H}, since
the WKB exactness is
sometimes declared as a generalization to the infinite dimensional case.
In order to
formulate quantity as path integral, there need to construct coherent
states over Grassmann
manifolds: we develop two ways. One is the algebraic method according to
the
Perelomov's prescription\cite{Perel} and the other is that in terms of the
Schwinger boson\cite{SB} as well as the canonical coherent state\cite{KS}.
These are the contents of section \ref{algcstrctn} and
\ref{scwcstrctn}, respectively. The path integral representation of the
character formula is
then given in  section \ref{piform} and the WKB approximation is performed
in section
\ref{wkbapprx}. The mechanism of the WKB exactness
is clarified in section \ref{bosepi} by making use of the Schwinger boson.
The final section
is devoted to related topics and remarks.
In Appendix A, proofs for
theorems, utilized in the section \ref{algcstrctn}, are given.
Finally in Appendix B a discussion on
the WKB exactness for the Feynman kernel is
given as a supplement; since which is considered to be more general than the
discussion in the text.

\section{The D-H formula on Grassmann manifolds}
\label{D-H}
In this section, we explain the validity of the D-H formula on
\(\grsmn{N}{n}\) to make a preparation for later discussions.

\subsection{Classical mechanics on \(\grsmn{N}{n}\)}
\label{cmgr}
Let \(\bbox{C}^{n}\) be the \(n\)-dimensional complex vector space. We
denote the space of \(m\times n\) matrices over \(\bbox{C}\) by
\(\Mnm{m}{n}\) and abbreviate \(\Mnm{n}{n}\) by \(\Mn{n}\).

We identify \(\grsmn{N}{n} \simeq\Un{N}/(\Un{n}\times\Un{N-n}) \) as a
phase space, assuming \(N\ge 2n\) for brevity's  sake,  to write
\be
\grsmn{N}{n}=\{P\in \Mn{N}|P^{2}=P,\
P^{\dagger}=P\
\mbox{ and }
\tr P=n\}\ .
\ee
\(P\) can be parameterized in terms of \(\xi\in\Mnm{N-n}{n}\) such that
\be
P=\pmatrix{
\dfrac{1}{\nxi}&
\dfrac{1}{\nxi} \xi^{\dagger}\cr
\xi\dfrac{1}{\nxi}&
\xi\dfrac{1}{\nxi}\xidag
}
=U(\xi)P_{1,\ldots,n}U^{\dagger}(\xi)\ ,
\label{grsmna}
\ee
where
\be
P_{1,\ldots,n} \equiv \pmatrix{
1_{n}&0\cr
0&0
} \ ,
\label{grsmn0}
\ee
and
\be
U(\xi) \equiv \pmatrix{
\dfrac{1}{\sqrt{\nxi}}&
- \dfrac{1}{\sqrt{\nxi}} \xi^{\dagger}\cr
\xi\dfrac{1}{\sqrt{\nxi}}&
\dfrac{1}{\sqrt{\mxi}}
}\ .
\ee
The parameterization \eqref{grsmna} cannot cover the whole phase space:
indeed
there exist other parameterizations such as
\be
(P_{\mu_{1},\ldots,\mu_{n}})_{\rho\lambda}
=\sum_{a=1}^{n}\delta_{\rho,\mu_{a}}\delta_{\lambda,\mu_{a}}, \quad 1 \leq
\mu_a
\leq N,
\label{grsmnb}
\ee
which tells us that there need \({N\choose n}\) kinds of local
parameterization. (Throughout the paper we use a convention for indices:
\(1 \leq \mu,\nu \leq N;  \  1 \leq a, b
\leq n;  \  n +1 \leq  i, j \leq N.\))
In order to obtain an
appropriate
parameterization in the neighborhood of
\(P_{\mu_{1},\ldots,\mu_{n}}\), we can utilize a unitary transformation
\(U(\mu_{1},\ldots,\mu_{n}|1,\ldots,n)\) satisfying
\be
U(\mu_{1},\ldots,\mu_{n}|1,\ldots,n)
P_{1,\ldots,n}
U^{\dagger}(\mu_{1},\ldots,\mu_{n}|1,\ldots,n)
=P_{\mu_{1},\ldots,\mu_{n}}\ .
\label{henkana}
\ee

The symplectic structure on \(\grsmn{N}{n}\) is defined through the
symplectic \(2\)-form
\be
\omega=i\tr(PdP\wedge dP)\ ,
\label{omegaa}
\ee
whose explicit form, under the above parameterization, is found as
\be
\omega=i\tr\{(\nxi)^{-1}d\xidag\wedge(\mxi)^{-1}d\xi\}
\label{omegab}
\ee
yielding to the \(\Un{N}\) invariant measure on \(\grsmn{N}{n}\),
\bea
&&\det\left[{
(\mxi)^{-1}\otimes\{(\nxi)^{-1}\}^{\cal T}
}\right](d\xi d\bxi)^{n(N-n)}
\nonumber\\
& = &
{1\over{\{\det(\nxi)\}^{N}}}(d\xi d\bxi)^{n(N-n)}\ ,
\label{invmes}
\eea
where the superscript \({\cal T}\) denotes the transpose of a matrix
and an abbreviation
\be
(dz d\bar{z})^{mn}
\equiv
\prod\limits_{\scriptstyle{1\le i\le m}\hfill\atop
\scriptstyle{1\le a\le n}\hfill}
d{\rm Re}(z_{ia})d{\rm Im}(z_{ia})\ ,
\label{abbmes}
\ee
has been employed. Our convention of tensor product is
\[
A\otimes B=
\pmatrix{
a_{11}B&a_{12}B&\cdots\cr
a_{21}B&a_{22}B&\cdots\cr
\vdots&\vdots&\ddots
},\
\text{ for }
A=(a_{ij}),\ B=(b_{ij})\ .
\]

Dynamical variables are defined through a linear mapping,
\be
X\in\Hn{N}\mapsto
F_{X}=\tr(PX)\in\boldkey{R},
\label{hamilta}
\ee
where \(\Hn{N}\) is the space of Hermitian matrices:
\be
\Hn{m}=\{X\vert X\in\Mn{m},\ X^{\dagger}=X\} \ .
\ee
The Poisson bracket is given, with the aid of \eqref{omegaa}, by
\be
\{F_{X},F_{Y}\}_{\mbox{\rm P.B.}}=\omega^{-1}(V_{X},V_{Y})
=F_{-i[X,Y]}\ ,
\label{poisson}
\ee
with \([X,Y]=XY-YX\),
where \(V_{X}\) is a vector field on \(\grsmn{N}{n}\) associated with
\(F_{X}\):
\[
V_{X}=\sum_{i,a}\left({
\delxi{F_{X}}{ia}\delxi{\ }{ia}
+\delbxi{F_{X}}{ia}\delbxi{\ }{ia}
}\right)\ .
\]
Note that the Poisson bracket \eqref{poisson} generates the \(u(N)\)
algebra.

An explicit form of the Hamiltonian function for a Hermitian matrix \(X\),
\be
X=\pmatrix{
A&B\cr
B^{\dagger}&D
},\
A\in\Hn{n},\
B\in\Mnm{n}{N-n},\
D\in\Hn{N-n},
\label{genham}
\ee
is read as
\begin{eqnarray}
F_{X}&=&\tr\{(\nxi)^{-1}\Phi_{X}\}\ ,
\nonumber\\
\Phi_{X}&=&
A+B\xi+\xidag B^{\dagger}+\xidag D\xi\ .
\label{hamiltb}
\end{eqnarray}

Introduce a \(1\)-form \(\t_{\kappa}\) such that
\be
d\t_{\kappa}=\omega,\quad
\t_{\kappa}=i\tr[\{\kappa\xidag d\xi-(1-\kappa)d\xidag\xi\}(\nxi)^{-1}]\ ;
\quad
\kappa\in\boldkey{R} \  ,
\ee
where the appearance of \(\kappa\) reflects an arbitrariness which does
not
affect
kinematics at all and is usually fixed by taking \( \kappa =1/2\).  An
action
for this
Hamiltonian system is thus found as
\bea
&&S=\int_{t_{1}}^{t_{2}}(\t_{\kappa}-F_{X}dt)
\label{action}\\
&=&\int_{t_{1}}^{t_{2}}dt\tr\left[{
(\nxi)^{-1}\left\{i(\kappa\xidag \dot\xi-(1-\kappa)\dot\xidag\xi)
-(A+B\xi+\xidag B^{\dagger}+\xidag D\xi)\right\}
}\right]\ . \nonumber
\eea
Equations of motion then are
\bea
\dot\xi&=&i(\xi A-D\xi-B^{\dagger}+\xi B\xi),
\nonumber\\
\dot\xidag&=&-i(A\xidag-\xidag D+\xidag B^{\dagger}\xidag-B).
\label{eqom}
\eea
If we put
\be
\exp(-iXt)=
\pmatrix{
\a(t)&\b(t)\cr
\gg(t)&\d(t)
}\ ,
\ee
we can find the solution:
\be
\xi(t)=\{\gg(t)+\d(t)\xi(0)\}\{\a(t)+\b(t)\xi(0)\}^{-1}\ .
\label{clsol1}
\ee
The solution \eqref{clsol1} takes the simplest form in the case of
block-diagonal
Hamiltonian given by \(B=0\):
\be
\xi(t)=U(t)\xi(0)V^{\dagger}(t),
\label{csol2}
\ee
where matrices \(V(t)\in \Un{n}\) and \(U(t)\in\Un{N-n}\) are given as
\be
V(t)=\exp(-iAt), \
U(t)=\exp(-iDt).
\label{unun}
\ee
In terms of \(\xi(t)\) \eqref{clsol1} the time dependence of \(P\)
\eqref{grsmna}
is read as
\be
P(t)=\pmatrix{
\dfrac{1}{1_{n}+\xi^{\dagger}(t)\xi(t)}&
\dfrac{1}{1_{n}+\xi^{\dagger}(t)\xi(t)} \xi^{\dagger}(t)\cr
\xi(t)\dfrac{1}{1_{n}+\xi^{\dagger}(t)\xi(t)}&
\xi(t)\dfrac{1}{1_{n}+\xi^{\dagger}(t)\xi(t)}\xidag(t)
}\ ,
\label{geodesic1}
\ee
so that
\be
P(t)=e^{-iXt}P(0)e^{iXt}\ .
\label{geodesic2}
\ee
Therefore we can recognize that the equations of motion \eqref{eqom}
describe the
action of
\(\Un{N}\) on \(\grsmn{N}{n}\)  in a local coordinate system.
We can thus see that classical mechanics on \(\grsmn{N}{n}\) formulated
above
 has a geometric interpretation.

As a final remark note that only
\(\xi(t)=0\) (\(0\le
t\le T\))  is allowed under the periodic boundary condition
\(\xi(T)=\xi(0)\)
for arbitrary \(T\) and \(X\).  In the subsequent section the same situation
can be
found in performing the WKB approximation.

\subsection{The D-H formula}
\label{D-Hform}
In this subsection we discuss the Duistermaat-Heckman
localization formula for the classical
system defined above. Start with a classical partition function
\be
{\cal Z}_{cl}(\b)=\int d\mu(\xi)\exp(-\b F_{H})\ ,\quad
\b>0\ ,
\label{partita}
\ee
where
\be
d\mu(\xi) \equiv {1\over{\{\det(\nxi)\}^{N}}} \left({d \xi d{\bar \xi} \over
\pi } \right)^{n(N-n)}
\ee
and satisfies
\be
\int d\mu(\xi)={\cal N}(n, N-n)
\label{measure1}
\ee
with
\be
{\cal N}(n, p) \equiv {{0!1!\cdots(n-1)!}\over {p!(p+1)!\cdots(p+n-1)!}},\
(p=0,1,2,\ldots)\ ,
\label{Nconst}
\ee
(whose verification is postponed until the next section: in \eqref{kanzenseia}
putting \(k\rightarrow 0\) we have \eqref{measure1}.)  Here \(F_{H}\) is a
Hamiltonian given in terms of a real diagonal matrix,
\be
H=\mbox{\rm diag}(h_{1},\ldots,h_{n},h_{n+1},\ldots,h_{N}),\quad
0<h_{1}<\cdots<h_{N},
\ee
to be found as
\bea
F_{H}=\tr\{(\nxi)^{-1}(H_{u}+\xidag H_{d}\xi)\}\ ,
\nonumber\\
H_{u}=\text{diag}(h_{1},\ldots,h_{n}),\
H_{d}=\text{diag}(h_{n+1},\ldots,h_{N})
\ .  \label{dhhamiltonian}
\eea

The first task toward the D-H formula is to search critical points of the
Hamiltonian by
solving
\be
\delxi{F_{H}}{ia}=0,\quad
\delbxi{F_{H}}{ia}=0,
\label{eqom2}
\ee
which coincide with the right hand sides of the equations of motion
\eqref{eqom}. In view of \eqref{partita} these critical points are nothing
but {\em saddle points\/}
of the integral.  For the present case, the saddle point conditions
\eqref{eqom2} become
\be
H_{u}\xidag-\xidag H_{d}=0,\
\xi H_{u}-H_{d}\xi=0. \label{eigen1}
\ee
According to the assumption, \(h_{1}<\cdots<h_{N}\), to the
Hamiltonian \eqref{dhhamiltonian}, the conditions cannot be met if
\(\xi\ne0\).

Thus we see that \(\xi=0\) is the only one solution of \eqref{eqom2}
{\em under this parameterization\/} of \(\grsmn{N}{n}\),
in other words, the unique fixed point in \(\Un{n}\times\Un{N-n}\)
(torus, \(\Un{1}^{N}\),  in this case) action \eqref{csol2}. Calculate
the  second derivative of \(F_{H}\) at \(\xi=0\); the Hessian
of the  Hamiltonian,
\be
\left.{
{\partial^{2}F_{H}\over
\partial\xi_{ia}\partial\bar{\xi}_{jb}}
}\right\vert_{\xi=0}
=
(H_{u})_{ab}\delta_{ji}-\delta_{ab}(H_{d})_{ji}\ .
\label{hessiancl}
\ee
Hence the contribution from this critical point to the saddle point
approximation of the integral \eqref{partita} is found to be
\bea
&&
\int\left({d\xi d\bxi\over\pi}\right)^{n(N-n)}
\exp\left\{{-\b\tr H_{u}-\b\tr(H_{d}\xi\xidag-H_{u}\xidag\xi)}\right\}
\nonumber\\
&=&{\exp(-\b\sum_{a=1}^{n}h_{a})\over{\b^{n(N-n)}}
\prod_{a=1}^{n}\prod_{i=n+1}^{N}
(h_{i}-h_{a})}\ .
\eea

As was stressed above we need
\({N\choose n}\)'s local parameterizations to
cover the whole phase space, therefore we must consider contributions from
other critical points.  To this end, recall the unitary transformation
given in \eqref{henkana}. A change of the local parameterization such
that
\be
P=U(\xi)P_{1,\ldots,n}U^{\dagger}(\xi)\mapsto
P_{U}=U(\mu_{1},\ldots,\mu_{n}|1,\ldots,n)P
U^{\dagger}(\mu_{1},\ldots,\mu_{n}|1,\ldots,n),
\ee
is equivalent to that of the Hermitian matrix in the Hamiltonian function
\be
F_{H}(P)=\tr(PH)\mapsto
F_{H}(P_{U})=F_{H^{\prime}}(P)=\tr(PH^{\prime}),
\ee
with
\be
H^{\prime}=U^{\dagger}(\mu_{1},\ldots,\mu_{n}|1,\ldots,n)H
U(\mu_{1},\ldots,\mu_{n}|1,\ldots, n) \ .
\ee
Once recognizing this, we can easily carry out the task; since
the new Hamiltonian after
the transformation is again diagonal without degeneracy so that a critical
point is always
located at \(\xi=0\) in each local parameterization. Summing up
those contributions, we obtain, as a result of the saddle point
approximation,
\be
{\cal Z}_{cl}(\b)\simeq
\sum_{\mu_{1}<\cdots<\mu_{n}}
{\exp(-\b\sum_{a=1}^{n}h_{\mu_{a}})\over{\b^{n(N-n)}}
\prod_{a=1}^{n}\prod_{\nu\in\bmu}
(h_{\nu}-h_{\mu_{a}})}\ ,
\label{clsadl}
\ee
where
\be
 \bmu\equiv\{1,\ldots,N\}\backslash\{\mu_{1},\ldots,\mu_{n}\} \ .
\label{barmu}
\ee
According to the D-H theorem the sum in the right hand side is now {\em
lifted to the exact
result.}

To see that this is true, that is, to convince the validity of the D-H
theorem, consider, instead of \eqref{partita}, the following expression:
\be
\int\limits_{\Hn{n}}
d\lambda
\int\limits_{\Mnm{N}{n}}
\left({{dzd\bar{z}}\over{\pi}}\right)^{Nn}
\exp\left[{
-\beta \tr (Z^{\dagger}HZ)
+i\tr\{\lambda(Z^{\dagger}Z-1)\}
}\right]\ ,
\label{partic}
\ee
where the integration domain of \(\lambda\) is \(\Hn{n}\) and new variables,
\bea
Z
& = &
 \pmatrix{
 Z_{u}\cr Z_{d} }
\ ,\quad  Z \in \Mnm{N}{n} \ , \nonumber \\
Z_{u} & = & \pmatrix{
z_{1,1}&\cdots&z_{1,n}\cr
\vdots&&\vdots\cr
z_{n,1}&\cdots&z_{n,n}
}
\ ,  \quad
Z_{d}=\pmatrix{
z_{n+1,1}&\cdots&z_{n+1,n}\cr
\vdots&&\vdots\cr
z_{N,1}&\cdots&z_{N,n}
}\ ,  \label{newvar}
\eea
have been introduced. An explicit form of \(\lambda\) is given as
\be
\lambda=
\pmatrix{
\lambda_{1}&\lambda_{1,2}&\cdots&\lambda_{1,n}\cr
\bar{\lambda}_{1,2}&\lambda_{2}&\ddots&\vdots\cr
\vdots&\ddots&\ddots&\lambda_{n-1,n}\cr
\bar{\lambda}_{1,n}&\cdots&\bar{\lambda}_{n-1,n}&\lambda_{n}
}\ ,
\label{lmata}
\ee
\bea
\lambda_{a} & \in & \bbox{R} \quad  (a=1,\ldots,n), \nonumber \\
\lambda_{a,b} & = &
{x_{a,b}-iy_{a,b}\over 2} ,  \quad  x_{a,b},y_{a,b}\in\bbox{R}, \quad
(1\le a<b\le n),
\label{lmatb}
\eea
so that the measure is
\be
d\lambda=\prod_{a=1}^{n}{d\lambda_{a}\over2\pi}
\prod_{a<b}{dx_{a,b}dy_{a,b}\over(2\pi)^{2}}.
\label{measher}
\ee

The \(\lambda\) integration gives delta functions which can be
solved by means of the following change of variables,
\be
Z=\pmatrix{1_{n}\cr \xi}
{1\over\sqrt{1_{n}+\xi^{\dagger}\xi}}\zeta; \quad \xi\in
M(N-n,n;\bbox{C}), \quad
\zeta\in M(n;\bbox{C}) \ ,
\label{henkan}
\ee
since
\be
Z^\dagger Z = \zeta^\dagger \zeta \ ,   \label{biline}
\ee
and the measure reads
\be
\left({dzd\bar{z}\over\pi}\right)^{Nn}=
\left({d\xi d\bar\xi\over\pi}\right)^{n(N-n)}
{1\over{\{\det(1_{n}+\xi^\dagger\xi)\}^{N}}}
\left({d\zeta d\bar\zeta\over\pi}\right)^{n^{2}}
\{\det(\zeta^{\dagger}\zeta)\}^{N-n}\ .
\label{chvarb}
\ee
(The integration with respect to \(\zeta\) is easily done convincing
that \eqref{partic} is equivalent to \eqref{partita}.) Therefore we can
recognize
that the role of
\(\lambda\) is to reduce the number of degrees of freedom from \(Nn\) to \(Nn
-n^2
=n(N-n)\). In this sense \(\lambda\) is called as multiplier and whose
coefficient as constraint\cite{SM}:
\be
\psi_{ab}\equiv (Z^{\dagger}Z)_{ab}-\delta_{ab}\approx 0  \ .
\label{constra}
\ee

However by exchanging the order of integrations the {\em Gaussian
integrations} with respect to
\(z_{\mu,a}\)'s result in
\be
\int\limits_{\Hn{n}}
d\lambda{\exp(-i\tr\lambda)\over\det(\b H\otimes
1_{n}-i1_{N}\otimes\lambda^{\cal T})}\ .
\label{partid}
\ee
With the help of the decomposition\cite{Mehta}
\be
\lambda=\Omega\lambda_{0} \Omega^{\dagger},\
\lambda_{0}=\text{diag}(l_{1},\ldots,l_{n}),\
\Omega\in\SUn{n},
\label{decompl}
\ee
we find after the \(\Omega\) integration
\be
\text{\eqref{partic}}=
\int_{-\infty}^{+\infty}{1\over n!}\prod_{a=1}^{n}{dl_{a}\over2\pi}
\prod_{a<b}(l_{a}-l_{b})^{2}
{\exp(-i\sum_{a=1}^{n}l_{a})\over
\prod_{a=1}^{n}\prod_{\mu=1}^{N}
(\b h_{\mu}-il_{a})}\ ,
\label{partie}
\ee
which leads to, by considering poles and zeros,
\be
\text{\eqref{partic}}=
\sum_{\mu_{1}<\cdots<\mu_{n}}
{\exp(-\b\sum_{a=1}^{n}h_{\mu_{a}})\over{\b^{n(N-n)}}
\prod_{a=1}^{n}\prod_{\nu\in\bmu}
(h_{\nu}-h_{\mu_{a}})}\ ,
\label{partif}
\ee
where again \(\bmu\) is given \eqref{barmu}. This is exactly the same
expression as
\eqref{clsadl}. Thus {\em the saddle point
approximation \eqref{clsadl}
itself contains the full information of the partition function} \({\cal
Z}_{cl}\).

We consider in what follows a quantum version of the D-H theorem,
that is, the WKB exactness in path integral. Our interpretation to the
D-H formula put here will be very helpful in the analysis.

\section{Coherent states of \(\Un{N}\) over \(\grsmn{N}{n}\)}
\label{cscstrctn}
In order to obtain the path integral representation we
construct coherent states of \(\Un{N}\) in this section. We first
consider the algebraic method proposed by Perelomov then explore an
generalization of the
Schwinger boson technique.

\subsection{Algebraic construction}
\label{algcstrctn}
The \(u(N)\) algebra in terms of generators \(E_{\mu\nu}\) reads
\be
[E_{\mu\nu},E_{\rho\sigma}]=
\delta_{\nu\rho}E_{\mu\sigma}-\delta_{\mu\sigma}E_{\rho\nu}, \quad
(1 \leq \mu, \nu, \rho, \sigma \leq N) .
\ee
First let us build up the coherent state in the fundamental
representation. The generators are given
\be
\left( E_{\mu\nu} \right)_{\rho\sigma} = \delta_{\mu \rho}\delta_{\nu
\sigma
}  \ .  \label{matrixrep}
\ee
Introduce an orthonormal set of basis vectors in \({\Bbb C}^{N}\),
\be
(\boldkey{e}_\mu)_{\nu}=\delta_{\mu\nu},
\qquad
\boldkey{e}_{\mu}^{\dagger} \boldkey{e}_\nu = \delta_{\mu \nu} \ ;
\label{orthonset}
\ee
to define a fiducial vector, by picking up first \(n\) \(\boldkey{e}_a\)
\((a=1, \cdots, n)\)
vectors out of \(N\) vectors, such that
\be
\vec{\cal E}_{N,n}={1\over\sqrt{n!}}
\sum_{\displaystyle\sigma\in{\cal S}_{n}}
\sgn{\sigma}\boldkey{e}_{\sigma(1)}\otimes\cdots\otimes\boldkey{e}_{\sigma(n)}
\in{\Bbb C}^{N^{n}}\ ,
\label{teata}
\ee
where  \({\cal S}_{n}\) denotes the symmetric group of order \(n\).
Consider the map
\be
\rho_{1}: GL(N;\bbox{C})\mapsto GL(N^{n};\bbox{C}),\
\rho_{1}(x)\equiv\otimes^{n}x
\left({
=\overbrace{x\otimes\cdots\otimes x}^{n}
}\right)\ ,
\label{defrho}
\ee
then it is obvious
\be
d\rho_{1}(\sum_{a=1}^{n}E_{aa}-\sum_{i=n+1}^{N}E_{ii})\vec{\cal E}_{N,n}=
n\vec{\cal E}_{N,n}\ ,
\label{defhwa}
\ee
\be
d\rho_{1}(E_{\mu i})\vec{\cal E}_{N,n}=
0\ ,
\label{defhwb}
\ee
where
\bea
d\rho_{1}(E)& \equiv &\left.{d\over{dt}}\right\vert_{t=0}
\rho_{1}(\exp tE)\nonumber\\
&=&\sum_{p=1}^{n}\otimes^{p-1}1_{N}\otimes E
\otimes^{n-p}1_{N}\ ,  \label{drho}
\eea
for \(E\in u(N)\). As for this fiducial
vector \(\vec{\cal
E}_{N,n}\), the following fact should be noted:
\begin{mylemma}
\be
\rho_{1}\left(B\right)
\vec{\Cal E}_{N,n}
=\det a\cdot\vec{\Cal E}_{N,n} \ ,
\label{detform}
\ee
where
\be
B\in GL(N;\bbox{C}),\quad
B=\pmatrix{
a&b\cr 0&c}  \ ,
\ee
with
\be
a\in GL(n;\Bbb C), \  b\in M(n,N-n;\Bbb C), \  c\in GL(N-n;\Bbb C)
\ .
\ee
\label{detformula}
\end{mylemma}
The proof is obvious so that be omitted. (However some comments would be
useful: what
this lemma signifies is that \(\det a\) (\(\vec{\Cal E}_{N,n}\)) is an
eigenvalue (eigenvector) of
\(\rho_{1}(B)\). If we put \(n=N\) in \eqref{detform}, the relation is nothing
but the definition of
the determinant.)

Let us now consider an element of \(\SUn{N}\) generated by an orthogonal
complement of the Lie algebra of \(\Un{n}\times\Un{N-n}\) ,
\be
S=\exp\pmatrix{
0&-\alpha^{\dagger}\cr
\alpha&0
}\ ; \quad  \a \in \Mnm{N-n}{n}\ ,
\label{unitaryopa}
\ee
which can be rewritten as
\be
S=
\pmatrix{
\dfrac{1}{\sqrt{\nxi}}&
- \dfrac{1}{\sqrt{\nxi}} \xi^{\dagger}\cr
\xi\dfrac{1}{\sqrt{\nxi}}&
\dfrac{1}{\sqrt{\mxi}}
}\ ; \quad  \xi\in\Mnm{N-n}{n} \ ,
\label{unitaryopb}
\ee
with
\be
\xi=\alpha{1\over\sqrt{\alpha^{\dagger}\alpha}}
\tan\sqrt{\alpha^{\dagger}\alpha}\ .
\ee
Noting the Gauss' decomposition \(S=LMU\),
with
\be
L=\pmatrix{
1_{n}&0\cr
\xi&1_{N-n}
},\
M=\pmatrix{
\dfrac{1}{\sqrt{\nxi}}&0\cr
0&\sqrt{\mxi}
},\
U=\pmatrix{
1_{n}&-\xidag\cr
0&1_{N-n}
},
\ee
we can obtain a desired (normalized) coherent state:
\bea
\vert\xi;1\rangle&\equiv &
\rho_{1}(LMU)\vec{\Cal E}_{N,n}\nonumber\\
&=&
\dfrac{1}{\{\det(1_{n}+\xi^{\dagger}\xi)\}^{1/2}}
\rho_{1}\left(L\right)
\vec{\Cal E}_{N,n}\ ,
\label{cs1}
\eea
where we have used the lemma \ref{detformula}.  While the unnormalized one
is given
by
\be
\vert\xi;1)\equiv
\rho_{1}\left(L\right)
\vec{\Cal E}_{N,n},\qquad
(\xi;1\vert\xi;1)=\det(1_{n}+\xi^{\dagger}\xi)  \ ,
\label{cs2}
\ee
whose second relation can be verified as follows: by noting that
\(
L\boldkey{e}_{a}
=\boldkey{e}_{a} + \sum_{i=n+1}^N
\xi_{ia}\boldkey{e}_{i}
\);
\(
( 1 \leq a \leq n)
\),
\be
\vert\xi;1)
=
{1\over\sqrt{n!}}
\sum_{\displaystyle\sigma\in{{\cal S}_{n}}}
\sgn{\sigma}
(\boldkey{e}_{\sigma(1)}+
\xi_{i_{1}\sigma(1)}\boldkey{e}_{i_{1}})
\otimes
\cdots
\otimes
(\boldkey{e}_{\sigma(n)}+
\xi_{i_{n}\sigma(n)}\boldkey{e}_{i_{n}})\ .
\label{csgena}
\ee
Here and for a while a repeated indices implies summation for brevity's
sake. Further
\be
(\boldkey{e}_{\sigma}^{\dagger}+
\bar\xi_{i\sigma}\boldkey{e}_{i}^{\dagger})
(\boldkey{e}_{\tau}+
\eta_{j\tau}\boldkey{e}_{j})
=\delta_{\sigma\tau}+(\xi^{\dagger}\eta)_{\sigma\tau} \ ,
\label{utilb}
\ee
then
\be
(\xi;1\vert\eta;1)=
\det(1_{n}+\xi^{\dagger}\eta)\ ,
\label{inpro1}
\ee
so that
\be
\langle\xi;1\vert\eta;1\rangle=
\det\left\{{
(1_{n}+\xi^{\dagger}\xi)^{-1/2}(1_{n}+\xi^{\dagger}\eta)
(1_{n}+\eta^{\dagger}\eta)^{-1/2}
}\right\}\ .
\label{csgenc}
\ee

Next calculate matrix elements of generators: in view of \eqref{drho} the
task is to estimate
\be
(\xi;1\vert d\rho_{1}(E_{\mu\nu})\vert\eta;1)
=\sum_{p=1}^{n}(\xi;1\vert
\otimes^{p-1}1_{N}\otimes E_{\mu\nu}
\otimes^{n-p}1_{N}\vert\eta;1)\ ,
\label{mtrxa}
\ee
which can be found,
after somewhat lengthy calculations, as
\bea
(\xi;1\vert d\rho_{1}(E_{ab})\vert\eta;1)
&=&
\det(1_{n}+\xi^{\dagger}\eta)\left({
{1\over{1_{n}+\xi^{\dagger}\eta}}
}\right)_{ba}\ ,
\label{mtrxab}\\
(\xi;1\vert d\rho_{1}(E_{ai})\vert\eta;1)
&=&
\det(1_{n}+\xi^{\dagger}\eta)\left({
\eta
{1\over{1_{n}+\xi^{\dagger}\eta}}
}\right)_{ia}\ ,
\label{mtrxai}\\
(\xi;1\vert d\rho_{1}(E_{ia})\vert\eta;1)
&=&
\det(1_{n}+\xi^{\dagger}\eta)\left({
{1\over{1_{n}+\xi^{\dagger}\eta}}\xi^{\dagger}
}\right)_{ai}\ ,
\label{mtrxia}\\
(\xi;1\vert d\rho_{1}(E_{ij})\vert\eta;1)
&=&
\det(1_{n}+\xi^{\dagger}\eta)\left({\eta
{1\over{1_{n}+\xi^{\dagger}\eta}}\xi^{\dagger}
}\right)_{ji}\ .
\label{mtrxij}
\eea

Armed with these, we obtain the matrix element of an arbitrary Hermitian
matrix,
\bea
H&=&\sum_{\mu,\nu}h_{\mu\nu}E_{\mu\nu}
\nonumber\\
&=&\sum_{a,b}h_{ab}E_{ab}
+\sum_{a,i}h_{ai}E_{ai}
+\sum_{j,b}h_{jb}E_{jb}
+\sum_{i,j}h_{ij}E_{ij}
\nonumber\\
&\equiv&\pmatrix{
H_{uu}&H_{ud}\cr
H_{du}&H_{dd}
}\ ,
\eea
such that
\be
(\xi;1\vert d\rho_{1}(H)\vert\eta;1)
=(\xi;1\vert\eta;1)\tr\left\{{
{1\over{1_{n}+\xi^{\dagger}\eta}}
(H_{uu}+H_{ud}\eta+\xi^{\dagger}H_{du}+\xi^{\dagger}H_{dd}\eta)
}\right\}\ ,
\label{mtrxh}
\ee
which is further rewritten to
\be
(\xi;1\vert d\rho_{1}(H)\vert\eta;1)=(\xi;1\vert\eta;1)
\tr\{P(\xi,\eta)H\}\ ,
\ee
or equivalently,
\be
\langle\xi;1\vert d\rho_{1}(H)\vert\eta;1\rangle
=\langle\xi;1\vert\eta;1\rangle\tr\{P(\xi,\eta)H\}\ ,
\label{matrxham}
\ee
where we have introduced a projection \(P(\xi,\eta)\),
\be
P(\xi,\eta)=
\pmatrix{
1_{n}\cr
\eta
}
(1_{n}+\xi^{\dagger}\eta)^{-1}
\pmatrix{
1_{n}&\xi^{\dagger}
}  \ ,
\label{defproj}
\ee
giving
\be
\tr\{P(\xi,\eta)H\}=\tr\left\{{
{1\over{1_{n}+\xi^{\dagger}\eta}}
(H_{uu}+H_{ud}\eta+\xi^{\dagger}H_{du}+\xi^{\dagger}H_{dd}\eta)
}\right\}\ .
\ee
(It should be noted that although the definition of \(P(\xi,\eta)\) looks
 singular in the domain
\(\{(\xi,\eta)\vert\det(1_{n}+\xi^{\dagger}\eta)=0\}\) there is no harm;
since the quantity
\(\tr\{P(\xi,\eta)H\}\) is always accompanied with
\(\langle\xi;1\vert\eta;1\rangle\) including
\(\det(1_{n}+\xi^{\dagger}\eta)\) in the numerator.)

Now we generalize the above result to a
higher dimensional representation. Consider a tensor product of the
coherent state
\be
\vert\xi;1\rangle\mapsto\vert\xi;k\rangle
\equiv\otimes^{k}\vert\xi;1\rangle\ ,\ k=0,1,2,\ldots\ ,
\label{csink}
\ee
as well as that of the representation
\be
\rho_{k}(x)\equiv\otimes^{k}(\rho_{1}(x)),\ x\in GL(N,\Bbb C)\ ,
\label{repink}
\ee
to put
\be
\vert\xi;k\rangle= \rho_{k}(LMU)\vec{\cal E}_{N,n}^{k}\ , \quad
\vec{\cal E}_{N,n}^{k}
\equiv
\otimes^{k}\vec{\cal E}_{N,n}\ .
\label{csinkb}
\ee
We designate this representation as the \(k\)-th representation. The
following relations are
obvious:
\be
d\rho_{k}(\sum_{a=1}^{n}E_{aa}-\sum_{i=n+1}^{N}E_{ii})\vec{\cal
E}_{N,n}^{k}=
kn\vec{\cal E}_{N,n}^{k}\ ,
\label{defhwka}
\ee
\be
d\rho_{k}(E_{\mu i})\vec{\cal E}_{N,n}^{k}=
0\ .
\label{defhwkb}
\ee
And
\begin{enumerate}
\rnc{\labelenumi}{\hbox to 2em{\hfill(\roman{enumi})\hfill}}
\item
\be
\langle\xi;k\vert\eta;k\rangle=\left[{
\det\left\{{
(1_{n}+\xi^{\dagger}\xi)^{-1/2}(1_{n}+\xi^{\dagger}\eta)
(1_{n}+\eta^{\dagger}\eta)^{-1/2}
}\right\}
}\right]^{k}\ ,
\label{inprk}
\ee
\item
\be
\langle\xi;k\vert d\rho_{k}(E_{\mu\nu})\vert\eta;k\rangle
= k\tr\{P(\xi,\eta)E_{\mu\nu}\}\langle\xi;k\vert\eta;k\rangle\ ,
\label{melka}
\ee
hence
\be
\langle\xi;k\vert d\rho_{k}(H)\vert\eta;k\rangle
= k\tr\{P(\xi,\eta)H\}\langle\xi;k\vert\eta;k\rangle \ .
\label{melha}
\ee
\end{enumerate}
In view of these relations, we notice the following facts: (i) the inner
product has a
form \(\{\langle\xi;1\vert\eta;1\rangle\}^{k}\) and (ii) the matrix
element
of a Hamiltonian is proportional to the parameter \(k\), which tells us
that the exponent of
path integral  is proportional to \(k\). Therefore we can perform the \(1 /
k\)-expansion
when \(k\) goes to large like the usual WKB-expansion with respect to \(
\hbar\).

If we declare that the state \(\vert\xi ;k\rangle\) is a coherent state, we
must check that the
resolution of unity does hold:
\be
{\bf 1}_{k}=\int d\mu(\xi;k)
\vert\xi;k\rangle\langle\xi;k\vert\ ,
\label{kanzensei}
\ee
with
\be
d\mu(\xi;k) \equiv {{\cal N}(n,k)\over{\cal N}(n,N-n+k)}
{1\over\{\det(1_{n}+\xi^{\dagger}\xi)\}^{N}}
\left({d\xi d{\bar \xi}\over\pi}\right)^{n(N-n)}
\ , \label{invmeas}
\ee
where \({\cal N}(n,k)\), has been given by \eqref{Nconst} and
\({\bf 1}_{k}\) is the
identity operator on the representation  space. To this end the following
formulae are
indispensable:
\begin{mytheorem}
Let \(dg\) be the normalized Haar measure on \(\Un{n}\).  Then for
\({}^{\forall}
p\in\boldkey{Z}_{+}\)  with \(\boldkey{Z}_{+}=\{0\}\cup\bbox{N}\) and
\({}^{\forall} X\in
M(n;\bbox{C})\),
there holds an integration formula
\be
\int_{\Un{n}}{dg\over (\det g)^{p}}\exp\left\{{\tr (gX)}\right\}
={\cal N}(n,p)|X|^{p},
\label{excellenta}
\ee
\be
{\cal N}(n,p)\equiv{{0!1!\cdots(n-1)!}\over {p!(p+1)!\cdots(p+n-1)!}}\ .
\label{nnp}
\ee
\label{excellent1}
\end{mytheorem}
\begin{mytheorem}
For  \({}^{\forall}X\in\Mn{n}\)
and \({}^{\forall}p \in\bbox{Z}_{+}\) there holds a differential formula
\be
\left\vert{\partial_{X}}\right\vert
\left\vert X\right\vert^{p}
=
p(p+1)\cdots(p+n-1)
\left\vert X\right\vert^{p-1}\ ,
\label{formulaa}
\ee
where \(|X|=\det X\)  and  \(\left\vert{\partial_{X}}\right\vert\) is
defined by
\be
\left\vert{\partial_{X}}\right\vert
\equiv \det\left({
\partial{\ }\over\partial x_{ij}
}\right)\ ,\
\text{ for }
X=(x_{ij})\ ,
\ee
which is valid even if \(p\) is negative integer
provided that \(X+X^{\dagger}\) is positive definite and \(|X|\ne 0\).
\label{excellent2}
\end{mytheorem}
Proofs of these formulae are straightforward but need a bit lengthy
calculations therefore we relegate them to Appendix A.  (Although the proof
of the theorem \ref{excellent2}, known as Cayley's formula, could be found
somewhere, for example, by using the Capelli's identity\cite{HW1}, we supply
our
own proof for a selfcontained purpose.)

Practically our target is to show instead of \eqref{kanzensei}
\be
(\alpha;k\vert\beta;k) = \int d\mu(\xi;k) \
(\alpha;k\vert\xi;k\rangle\langle\xi;k\vert\beta;k) \  ;  \quad
\alpha,\beta\in
M(N-n,n;\bbox{C}) \ ,
\label{kanzenseib}
\ee
that is,
\bea
 \{\det(1_{n}+\alpha^{\dagger}\beta)\}^{k}
&=& \int d\mu(\xi;k) \ {\{\det(1_{n}+\alpha^{\dagger}\xi)\}^{k}
\{\det(1_{n}+\xi^{\dagger}\beta)\}^{k}
\over\{\det(1_{n}+\xi^{\dagger}\xi)\}^{k}} \nonumber \\
& = & {{\cal N}(n,k)\over{\cal N}(n,N-n+k)}
\int \left({d\xi d{\bar \xi}\over\pi}\right)^{n(N-n)}
{1\over\{\det(1_{n}+\xi^{\dagger}\xi)\}^{N}} \nonumber \\
 & \times &
{\{\det(1_{n}+\alpha^{\dagger}\xi)\}^{k}
\{\det(1_{n}+\xi^{\dagger}\beta)\}^{k}
\over\{\det(1_{n}+\xi^{\dagger}\xi)\}^{k}} \ .
\label{kanzenseia}
\eea
Establishing \eqref{kanzenseia} is equal to establishing
\eqref{kanzensei}; since these
relations hold for any
\(\vert\a;k)\) and
\(\vert\b;k)\). In order to accomplish this, there need two
others relations.
\begin{mycorol}
\label{coro1}
For \( {}^{\forall} k\in\boldkey{Z}_{+}\), there holds a formula for
Gaussian type
integration  over
\(\Mnm{m+n}{n}\):
\be
\int\left(dzd\bar{z}\over\pi\right)^{n(m+n)}
\vert Z^{\dagger}Z\vert^{k}
\exp\left\{{-\tr (Z^{\dagger}Z)}\right\}
=
{{\cal N}(n,m)\over{\cal N}(n,m+k)}\ .
\label{gausscor1}
\ee
\end{mycorol}
{\em Proof.\/} By making use of an identity,
\be
\vert Z^{\dagger}Z\vert^{k}\exp\left\{{-\tr (Z^{\dagger}Z)}\right\}
=\left.{
\vphantom{{\partial\over\partial X}}
(-1)^{nk}|\partial_{X}|^{k}
}\right\vert_{X=1_{n}}
\exp\left\{{-\tr (XZ^{\dagger}Z)}\right\}\ ,
\label{trick0}
\ee
the left hand side of \eqref{gausscor1} is rewritten as
\bea
&&
\left.{
\vphantom{{\partial\over\partial X}}
(-1)^{nk}|\partial_{X}|^{k}
}\right\vert_{X=1_{n}}
\int\left(dzd\bar{z}\over\pi\right)^{n(m+n)}
\exp\left\{{-\tr (XZ^{\dagger}Z)}\right\}
\nonumber\\
&=&
\left.{
\vphantom{{\partial\over\partial X}}
(-1)^{nk}|\partial_{X}|^{k}}\right\vert_{X=1_{n}}
\vert X\vert^{-(m+n)}\ ,
\label{gausscor2}
\eea
which becomes, by a repeated use of the formula \ref{excellent2},
\bea
\text{\eqref{gausscor2}}
&=&{(m+n)!\over m!}\times{(m+n+1)!\over (m+1)!}\times\cdots\times
{(m+n+k-1)!\over(m+k-1)!}
\nonumber\\
&=&
\dfrac{{\cal N}(n,m)}{{\cal N}(n,m+k)}\ .
\label{gausscor3}
\eea
\begin{mycorol}
For \( {}^{\forall}p,q\in\boldkey{Z}_{+}\) and \(
{}^{\forall}A,B\in\Mnm{m+n}{n}\),
there holds
\bea
&&\int_{\Un{n}}{dg\over(\det g)^{p}}
\int\left(dzd\bar{z}\over\pi\right)^{n(m+n)}
\vert Z^{\dagger}Z\vert^{q}
\exp\{-\tr (Z^{\dagger}Z-Z^{\dagger}A-gB^{\dagger}Z)\}
\nonumber\\
&=&{{\cal N}(n,p){\cal N}(n,m+p)\over{\cal N}(n,m+p+q)}
\vert B^{\dagger}A\vert^{p}\ .
\label{eqcoro21}
\eea
\label{coro2}
\end{mycorol}
{\em Proof.\/} Integrate with respect to \(Z,Z^{\dagger}\) and follow a
similar procedure as
above to find
\be
\text{\eqref{eqcoro21}}=
\left.{
\vphantom{{\partial\over\partial X}}
(-1)^{nq}|\partial_{X}|^{q}}\right\vert_{X=1_{n}}
\int_{\Un{n}}{dg\over(\det g)^{p}}
\exp\{\tr(gB^{\dagger}AX^{-1})\}
\vert X\vert^{-(m+n)}\ .
\label{eqcoro22}
\ee
The formula \ref{excellent1} enables us to perform the \(g\) integration
giving
\be
\text{\eqref{eqcoro22}}=
\left.{
\vphantom{{\partial\over\partial X}}
(-1)^{nq}|\partial_{X}|^{q}}\right\vert_{X=1_{n}}
\vert X\vert^{-(m+n+p)}
{\cal N}(n,p)\vert B^{\dagger}A\vert^{p} \ ,
\label{eqcoro23}
\ee
which turns out, again by the repeated use of the formula
\ref{excellent2}, to be the right
hand side of
\eqref{eqcoro21}.

Now we can proceed to our target: first rewrite the left hand side of
\eqref{kanzenseia} by use of the  formula \ref{excellent1} as
\be
\{\det(1_{n}+\alpha^{\dagger}\beta)\}^{k}
=
{1\over{\cal N}(n,k)}\int_{\Un{n}}{dg\over(\det g)^{k}}
\exp\left[{\tr \left\{{g(1_{n}+\alpha^{\dagger}\beta)}\right\}}\right]\ ,
\label{subarasiia}
\ee
whose integrand can be expressed by the Gaussian
integration with respect to \(Z\in\Mnm{N}{n}\),
\bea
&&
\exp\left[{\tr \left\{{g(1_{n}+\alpha^{\dagger}\beta)}\right\}}\right]
\nonumber\\
&=&
\int\left({dzd{\bar z}\over\pi}\right)^{Nn}
\exp\left[{
-\tr\left\{{
Z^{\dagger}Z-Z^{\dagger}\pmatrix{1_{n}\cr \beta}
-(1_{n},\alpha^{\dagger})Zg
}\right\}
}\right]\ .
\label{trick1}
\eea
A change of variables, \(Z \rightarrow ( \xi , \zeta)\), as in the previous
section from \eqref{henkan} to \eqref{biline}, gives
\begin{eqnarray}
\text{\eqref{trick1}}&=&\int\left({d\xi d{\bar
\xi}\over\pi}\right)^{n(N-n)}
{1\over \{\det(1_{n}+\xi^{\dagger}\xi)\}^{N}}
\int\left({d\zeta d{\bar \zeta}\over\pi}\right)^{n^{2}}
(\det \zeta^{\dagger}\zeta)^{N-n}
\nonumber\\
& &
\times
\exp\left[{
-\tr\left\{{
\zeta^{\dagger}\zeta-\zeta^{\dagger}
(1_{n}+\xi^{\dagger}\xi)^{-1/2}
(1_{n}+\xi^{\dagger}\beta)
}\right.}\right.
\nonumber\\
& &
\mathop{\hphantom{
\times
\exp(-\tr\zeta^{\dagger}\zeta)
}}
\left.{\left.{
-g(1_{n}+\alpha^{\dagger}\xi)
(1_{n}+\xi^{\dagger}\xi)^{-1/2}\zeta
}\right\}
}\right]\ .
\label{trick2}
\end{eqnarray}
Substituting \eqref{trick2} into \eqref{subarasiia} then integrating with
respect to \(\zeta\)
and \(g\) with the aid of the formula \ref{coro2}, we find
\begin{eqnarray}
&&\text{r.h.s of \eqref{subarasiia}}=
{{\cal N}(n,k)\over{\cal N}(n,N-n+k)}
\label{trick5}\\
&&\times\int\left({d\xi d{\bar \xi}\over\pi}\right)^{n(N-n)}
{1\over\{\det(1_{n}+\xi^{\dagger}\xi)\}^{N}}
{\{\det(1_{n}+\alpha^{\dagger}\xi)\}^{k}
\{\det(1_{n}+\xi^{\dagger}\beta)\}^{k}
\over\{\det(1_{n}+\xi^{\dagger}\xi)\}^{k}}\ ,
\nonumber
\end{eqnarray}
which is nothing but the right hand side of \eqref{kanzenseia}.

We now consider another version of coherent state
with the aid of the Schwinger boson technique before going into a path
integral discussion.

\subsection{Coherent state via Schwinger boson}
\label{scwcstrctn}
As was stressed in section \ref{D-Hform}, essence of the classical D-H
theorem can
easily be grasped by increasing degrees of freedom while balancing them with
the aid of the  multiplier
\(\lambda\): we call such a view point as that of constrained system. If we
could find a similar
way in a quantum case, then establishment of the localization would be
obvious. Fortunately
we know such a candidate which might realize our expectation: the method of
Schwinger boson.
There in order to obtain a group representation generators of a group are
expressed by
creation and annihilation operators.  The representation space is thus the
Fock space whose dimension is too big for a group (especially for a compact
one.)
Therefore there needs
some constraint to reduce the whole space. In a  simple case such
as \(\bbox{C}P^{N}\)\cite{FFKS1,FFKS2} it is realized by fixing the total
particle number.
In this way, the scenario would be hopeful.

Consider operators,
\begin{equation}
a=
\pmatrix{
a_{u}\cr a_{d}
}
\ ,\
a_{u}=\pmatrix{
a_{1,1}&\cdots&a_{1,n}\cr
\vdots&&\vdots\cr
a_{n,1}&\cdots&a_{n,n}
}
\ ,\
a_{d}=\pmatrix{
a_{n+1,1}&\cdots&a_{n+1,n}\cr
\vdots&&\vdots\cr
a_{N,1}&\cdots&a_{N,n}
}\ ,
\end{equation}
which obey (\(1 \leq \mu, \nu \leq N; \ 1 \leq a, b \leq n\))
\be
[a_{\mu,a},a_{\nu,b}^{\dagger}]=\delta_{\mu\nu}\delta_{ab} \ ,   \
[a_{\mu,a},a_{\nu,b}] = [a_{\mu,a}^{\dagger}, a_{\nu,b}^{\dagger}] =0 \
.   \label{ccr}
\ee
In terms of these operators \(u(N)\) generators are realized:
\be
\hat{E}_{\mu\nu}=\tr(a^{\dagger}E_{\mu\nu}a)\ ;
\label{Emunu}
\ee
\be
[\hat{E}_{\mu\nu},\hat{E}_{\rho\sigma}]
=(\delta_{\nu\rho}\hat{E}_{\mu\sigma}
-\delta_{\sigma\mu}\hat{E}_{\rho\nu})\ .
\label{UNalg}
\ee
The Fock space  \(\cal F\) is designated as
\be
{\cal F}=\text{Span}\{\vert(n_{\mu,a})\rangle
=\prod_{\mu,a}{1\over\sqrt{n_{\mu,a}!}}a^{\dagger}_{\mu,a}
{}^{n_{\mu,a}}
\vert 0\rangle\ ,\ a_{\mu,a}\vert 0\rangle=0,\ n_{\mu,a}\in
\bbox{Z}_{+}\}\ .
\label{fock}
\ee

Introduce a usual canonical coherent state\cite{KS}:
\begin{equation}
\vert Z\rangle
\equiv\exp\left\{{
\tr(a^{\dagger}Z-Z^{\dagger}a)
}\right\}
\vert 0\rangle
=\exp\left\{-{1\over2}\tr(Z^{\dagger}Z)\right\}
\exp\left\{\tr (a^{\dagger}Z)\right\}\vert 0\rangle\ ,
\label{bosecs}
\end{equation}
\be
{\bf 1}=\int\left({dzd\bar{z}\over\pi}\right)^{Nn}
\vert Z\rangle\langle Z\vert,
\label{bosekanzen}
\ee
\be
\langle Z\vert Z^{\prime}\rangle
=\exp\left\{{-{1\over2}\tr(Z^{\dagger}Z+Z^{\prime\dagger}Z^{\prime})
+\tr(Z^{\dagger}Z^{\prime})}\right\}\ ,
\label{boseinpr}
\ee
where \({\bf 1}\) denotes the identity operator on \(\cal F\) and \(Z\) has
been given by
\eqref{newvar}.

Let us consider a Hermitian projection operator,
\be
P_{k}\equiv
\int\left({
dzd{\bar z}\over\pi
}\right)^{Nn}
\int_{\Un{n}}{dg\over(\det g)^{k}}
\vert Zg\rangle
\langle Z \vert\ .
\label{projection}
\ee
A simple inspection leads to
\be
P_{k}P_{k^{\prime}}=P_{k}\delta_{k,k^{\prime}} \ , \qquad
P_{k}^{\dagger}=P_{k}\ .
\ee
In what follows we can recognize that this projection operator indeed
reduces \(\cal F\) to
the space of the \(k\)-th representation in the previous section, but before
proceeding it is
instructive to discuss how to find the form of \(P_k\) as
\eqref{projection}. By noting
\be
\exp\{i\tr(\lambda a^{\dagger}a)\}\vert Z\rangle=
\vert Zg\rangle,\quad
g=e^{i\lambda}\in \Un{n}\  ,
\label{unaction}
\ee
so that
\be
{1\over(\det g)^{k}} \vert Zg\rangle
= \exp\left[ i\tr\left\{ \lambda
( a^{\dagger}a- k
) \right\} \right]  \vert Z \rangle \ ,
\ee
which immediately reminds us of the multiplier part of \eqref{partic} by
replacing
\(a(a^\dagger)\) with \(Z(Z^\dagger)\). (However note
that the  difference of the
integration domain of \(\lambda\); in
\eqref{partic} it is infinite but in \eqref{projection} it is bounded. We
consider more on this issue in section 5.)

The trace of \(P_k\) can be calculated as
\begin{eqnarray}
\Tr P_{k}& = &
\int\left({
dzd{\bar z}\over\pi
}\right)^{Nn}
\int_{\Un{n}}{dg\over(\det g)^{k}}
\langle Z \vert
Zg\rangle
\nonumber\\
& = &
{\cal N}(n,k)\int\left({
dzd{\bar z}\over\pi
}\right)^{Nn}
|Z^{\dagger}Z|^{k}
\exp\left\{{-\tr (Z^{\dagger}Z)}\right\}
\nonumber\\
& = &
{{\cal N}(n,k){\cal N}(n,N-n)
\over{\cal N}(n,N-n+k)}\ ,
\label{dimension}
\end{eqnarray}
where we have used \eqref{boseinpr} then the formula
\ref{coro1}. (We have employed the notation \(\Tr(\cdots)\) for the trace
over all the Fock
space while \(\Tr_{k}(\cdots)\) used below for that within the \(k\)-th
representation  space and those should be distinguished from
\(\tr(\cdots)\) used for matrix valued quantities.)
\eqref{dimension} implies the dimension of the \(k\)-th representation,
that is, the number of independent vectors in \({\cal F}\) which
satisfy \(n^{2}\) physical state conditions:
\be
(a^{\dagger}a)_{a,b}\vert\text{phys}\rangle
=k\delta_{a,b}\vert\text{phys}\rangle\ .
\label{physcond}
\ee
In partcular, for the case of \(k=1\),
the relation \eqref{dimension}
implies an arbitrariness of choosing a fiducial vector
\eqref{teata}.
It is symmetric with respect to \(n\) and
\(N-n\), which clearly reflects the  nature of the base manifold
\(\grsmn{N}{n}\), namely \(\grsmn{N}{n}\cong\grsmn{N}{N-n}\) and is easily
checked by an explicit calculation:
\be
{{\cal N}(n,k){\cal N}(n,N-n)
\over{\cal N}(n,N-n+k)}
=
{{\cal N}(N-n,k){\cal N}(N-n,n)
\over{\cal N}(N-n,n+k)} \ .
\label{dim2}
\ee

Now we show that
\be
P_{k}=\int d\mu(\xi;k)
\vert\xi;k\rangle\langle\xi;k\vert\ ,
\label{kanzen3}
\ee
where \(\vert\xi;k\rangle\) is the coherent state of \(\Un{N}\) over
\(\grsmn{N}{n}\) derived previously but now given by
\be
\vert\xi;k\rangle \equiv
{1\over
\{\det(1_{n}+\xi^{\dagger}\xi)\}^{k/2}}
\exp\{\tr(a^{\dagger}_{d}\xi a_{u})\}
\sqrt{ {\cal N}(n,k) }(\det a_{u}^{\dagger})^{k}\vert0\rangle \ .
\label{csosc1}
\ee
Thus we can regard \eqref{kanzen3} as the resolution of
unity
\eqref{kanzensei}.

In order to reach the resolution of unity \eqref{kanzen3} and the coherent
state
\eqref{csosc1}, first rewrite \eqref{projection} to
\be
{P}_{k}
=
\int\left({
dzd{\bar z}\over\pi
}\right)^{Nn}
\int_{\Un{n}}{dg_{1}dg_{2}\over\{\det (g_{1}g_{2})\}^{k}}
\vert Zg_{1}\rangle\langle Zg_{2}^\dagger \vert  \ ,
\label{henkei0}
\ee
which can be recognized directly by putting \(Z g_2^\dagger \mapsto Z\)
and  \(g_1g_2 \mapsto g\) and finally performing the trivial integration
with respect to  \(g_2\).
Then note that from \eqref{boseinpr},
\bea
\vert Z g \rangle
& = &
\exp \left\{ - {1 \over 2} \tr (Z^\dagger Z) \right\}
\exp\left\{{
\tr(a^{\dagger}Zg)
}\right\}\vert 0\rangle  \nonumber \\
& = &
\exp \left\{ - {1 \over 2} \tr (Z^\dagger Z) \right\}
\exp\left[{
\tr\{(a_{u}^{\dagger}Z_{u}+a_{d}^{\dagger}Z_{d})g\}
}\right]\vert 0\rangle
\nonumber\\
&=&
\exp \left\{ - {1 \over 2} \tr (\zeta^\dagger \zeta)  \right\}
\exp\left\{{
\tr(a_{d}^{\dagger}\xi a_{u})
}\right\}
\exp\left\{{
\tr(a_{u}^{\dagger}\Lambda^{-1/2}\zeta g)
}\right\}\vert 0\rangle\ ,
\label{henkei1}
\eea
with \(\Lambda=1_{n}+\xi^{\dagger}\xi\), where use has been made of
the change of variables \eqref{henkan} and the
Campbell-Baker-Hausdorff formula to the final expression. (As was
discussed in the previous section,
employing \eqref{henkan} is nothing but choosing some fiducial vector.)
By means of \eqref{henkei1} and  the formula \ref{excellent1} we find
\bea
&&\int_{\Un{n}}{dg_{1}\over(\det g_{1})^{k}}
\vert Zg_{1}\rangle
\nonumber\\
&=&
{\cal N}(n,k)
\exp \left\{ - {1 \over 2} \tr (\zeta^\dagger \zeta)  \right\}
\exp\left\{{
\tr(a_{d}^{\dagger}\xi a_{u})
}\right\}
\{\det (a_{u}^{\dagger}\Lambda^{-1/2}\zeta)\}^{k}
\vert 0\rangle\ .
\label{ginteg}
\eea
Also note the relation, obtained from the
formula \ref{coro1},
\be
\int\left({d\zeta d{\bar \zeta}\over\pi}\right)^{n^{2}}
|\zeta^{\dagger}\zeta|^{N-n+k}
\exp \left\{ -\tr (\zeta^\dagger \zeta)  \right\}
=
{1\over{\cal N}({n},{N-n+k})}\ .
\label{henkei3}
\ee
Substituting \eqref{ginteg} (and whose conjugate) into \eqref{henkei0}
then utilizing
\eqref{henkei3} , we finally arrive at
\bea
P_{k}&=&
{{{\cal N}({n},{k})}^{2}\over{\cal N}(n,N-n+k)}
\int\left({d\xi d{\bar \xi}\over\pi}\right)^{n(N-n)}
{1\over\{\det(\nxi)\}^{N+k}}
\nonumber\\
&&\times
\exp\left\{{
\tr(a_{d}^{\dagger}\xi a_{u})
}\right\}
(\det a_{u}^{\dagger})^{k}
\vert 0\rangle
\langle 0\vert
(\det a_{u})^{k}
\exp\left\{{
\tr(a_{u}^{\dagger}\xi^{\dagger} a_{d})
}\right\}\ .
\label{henkei2}
\eea
Therefore we have established \eqref{kanzen3} and \eqref{csosc1}.
By comparing \eqref{csosc1} with \eqref{cs1} and \eqref{csink} the state
\(\sqrt{{\cal
N}(n,k)} (\det a_{u}^{\dagger})^{k}\vert0\rangle\) should be identified to
\(\vec{\cal E}_{N,n}^{k}\). Therefore the projection
operator onto the subspace can now be regarded as the resolution
of
unity in the space of the \(k\)-th representation.

\section{Path integral and WKB}
\label{piofcf}

In this section we first build up a path integral representation of a
character formula by
means of the coherent states developed previously. We then perform the WKB
approximation to find the result is exact. The mechanism of the exactness
is uncovered by
reformulating the theory in terms of the generalized Schwinger boson
technique.

\subsection{Path integral representation}
\label{piform}
Take a Hamiltonian in the \(k\)-th representation
\be
\hat{H}=d\rho_{k}(H)\ ,\
H={\rm diag}(h_{1},\ldots,h_{N}),\ h_{\mu}\in\bbox{R},\
h_{1}<\cdots<h_{N} \ .
\label{hamiltn}
\ee
Consider the trace of the time evolution operator which is from now on
designated as the
{\em character formula} of the
\(k\)-th representation:
\bea
{\cal Z}_{k}(T)
&\equiv&\Tr_{k}\rho_{k}(e^{-iHT})
\nonumber\\
&=&
\lim\limits_{M\rightarrow\infty}
\int d\mu(\xi;k)\langle\xi;k\vert
\left\{{
d\rho_{k}(1_{N}-i\Delta t H)
}\right\}^{M}
\vert\xi;k\rangle\ ,
\label{characa}
\eea
where \(\Delta t=T/M\).
Inserting the resolution of unity \eqref{kanzensei} into the final
expression successively, we
obtain
\be
{\cal Z}_{k}(T)=
\lim\limits_{M\rightarrow\infty}
\int\limits_{\text{PBC}} \prod_{j=1}^{M}d\mu(\xi(j);k)\langle\xi(j);k\vert
d\rho_{k}(1_{N}-i\Delta t H)
\vert\xi(j-1);k\rangle\ ,
\label{characb}
\ee
where as before ``PBC'' means \(\xi(0)=\xi(M)\). In view of \eqref{melha},
\bea
&&\langle\xi(j);k\vert d\rho_{k}(1_{N}-i\Delta t H)\vert\xi(j-1);k\rangle
\nonumber\\
&=&\langle\xi(j);k\vert\xi(j-1);k\rangle
\left[{1-ik\Delta t\tr\{P(\xi(j),\xi(j-1))H\}}\right]
\nonumber\\
&=&
\langle\xi(j);k\vert\xi(j-1);k\rangle
\exp\left[{-ik\Delta t\tr\{P(\xi(j),\xi(j-1))H\}}\right]
\nonumber\\
&&
\times
\left\{{1+O((\Delta t)^{2})}\right\} \ .
\label{characc}
\eea
Employing the expression \eqref{inprk} for
\(\langle\xi(j);k\vert\xi(j-1);k\rangle\), we obtain
\begin{eqnarray}
{\cal Z}_{k}(T)
& = &
\lim\limits_{M\rightarrow\infty}
\int\limits_{\text{PBC}} \prod_{j=1}^{M}d\mu(\xi(j);k)
\nonumber\\
&&
\hphantom{
\lim\limits_{M\rightarrow\infty}
}
\times
\exp\left[{
-k\sum_{i=1}^{M}\tr
\left\{{
\log(1_{n}+\xi^{\dagger}(i)\xi(i))
-\log(1_{n}+\xi^{\dagger}(i)\xi(i-1))
}\right\}
}\right]
\nonumber\\
& &
\hphantom{
\lim\limits_{M\rightarrow\infty}
}
\times
\exp\left[{
-ik\Delta t
\sum_{j=1}^{M}
\tr\{P(\xi(j),\xi(j-1))H\}
}\right]\ ,
\label{characf}
\end{eqnarray}
where we have discarded terms of \(O((\Delta t)^{2})\), whose fact also
brings us to
\begin{eqnarray}
{\cal Z}_{k}(T)& = &
(\det V(T))^{k}
\lim\limits_{M\rightarrow\infty}
\int\limits_{\text{PBC}} \prod_{j=1}^{M}d\mu(\xi(j);k)
\nonumber\\
&&
\times
\exp\left[{
-k\sum_{i=1}^{M}\tr
\left\{{
\vphantom{
\sum_{i=1}^{M}\tr
\left\{{
\log(1_{n}+\xi^{\dagger}(i)\xi(i))}\right\}
}
\log(1_{n}+\xi^{\dagger}(i)\xi(i))
}\right.}\right.
\nonumber\\
&&
\left.{\left.{
\vphantom{
\sum_{i=1}^{M}\tr
\left\{{
\log(1_{n}+\xi^{\dagger}(i)\xi(i))}\right\}
}
\hphantom{
(\det V(T))^{k}
}
-\log(1_{n}+\xi^{\dagger}(i)U(\Delta t)\xi(i-1)V^{\dagger}(\Delta	t))
}\right\}
}\right]\ ,
\label{characg}
\end{eqnarray}
where \(U(t)\) and \(V(t)\) have been defined by \eqref{unun},
\be
U(t)=e^{-iH_{d}t}\in \Un{N-n}\ ,\
V(t)=e^{-iH_{u}t}\in \Un{n}\ ,
\label{ununa}
\ee
\be
H_{u}=\text{diag}(h_{1},\ldots,h_{n})\ ,\
H_{d}=\text{diag}(h_{n+1},\ldots,h_{N})\ .
\ee
By noting that
\bea
&&\exp\left[{-k\tr\left\{{\log(1_{n}+\xi^{\dagger}(i)\xi(i))
-\log(1_{n}+\xi^{\dagger}(i)U(\Delta t)\xi(i-1)V^{\dagger}(\Delta t))
}\right\}}\right]
\nonumber\\
&=&
\left\{{\det(1_{n}+\xi^{\dagger}(i)U(\Delta t)\xi(i-1)
V^{\dagger}(\Delta t))\over\det(1_{n}+\xi^{\dagger}(i)\xi(i))}\right\}^{k}
  \ ,
\eea
\eqref{characg} becomes
\bea
{\cal Z}_{k}(T)&=&
(\det V(T))^{k}
\lim\limits_{M\rightarrow\infty}
\int\limits_{\text{PBC}} \prod_{i=1}^{M}d\mu(\xi(i);k)
\nonumber\\
&&
\hphantom{(\det V(T))^{k}}
\times
\prod_{j=1}^{M}
\langle \xi(j)V(\Delta t);k\vert U(\Delta t)\xi(j-1);k\rangle\ ,
\label{characga}
\eea
where use has been made of \eqref{inprk}. The multiple integration can be
carried out
with the aid of
\eqref{kanzenseia} to yield
\be
{\cal Z}_{k}(T)=
(\det V(T))^{k}
\int d\mu(\xi;k)
\left\{
{\det(1_{n}+\xi^{\dagger}U(T)\xi V^{\dagger}(T))
\over
\det(1_{n}+\xi^{\dagger}\xi)}
\right\}^{k}\ .
\label{characgb}
\ee
(See a further discussion in Appendix \ref{Feynman}.)
By means of of the formula \ref{excellent1} the final integration can be
performed, however the task is postponed for a while.

\subsection{The WKB approximation}
\label{wkbapprx}
Let us examine the WKB approximation of the path integral expression
\eqref{characg}. Equations of motion are
\bea
&&
\xi(j)\{1_{n}+\xi^{\dagger}(j)\xi(j)\}^{-1}\nonumber\\
&= & U(\Delta t)\xi(j-1)V^{\dagger}(\Delta t)
\{1_{n}+\xi^{\dagger}(j)U(\Delta t)\xi(j-1)V^{\dagger}(\Delta t)\}^{-1}\ ,
\label{eqofmotiona}
\eea
\bea
&&
\{1_{n}+\xi^{\dagger}(j)\xi(j)\}^{-1}\xi^{\dagger}(j)\nonumber\\
&= & \{1_{n}+V^{\dagger}(\Delta t)\xi^{\dagger}(j+1)U(\Delta
t)\xi(j)\}^{-1}
V^{\dagger}(\Delta t)\xi^{\dagger}(j+1)U(\Delta t) \ .
\label{eqofmotionb}
\eea
Under the present situation, that is, calculating the character, solutions
should meet
the periodic  boundary condition \(\xi(0)=\xi(M)\).
Clearly, only \(\xi(j)=0\) for all \(1\le j\le M\)
can fulfill the condition. Therefore, by putting \(\xi=z/\sqrt{k}\) and
noting
\be
{{\cal N}(n,k)\over{\cal N}(n,N-n+k)}
\stackrel{k\rightarrow\infty}{\sim} k^{n(N-n)}   \ ,  \label{largek}
\ee
the dominant contribution from this classical solution is read as
\begin{eqnarray}
{\tilde{\cal Z}}_{k}(T)
& \stackrel{k \rightarrow \infty}{\sim} &
\lim\limits_{M\rightarrow\infty}
\int\limits_{\text{PBC}} \prod_{j=1}^{M}\left({dz(j)d{\bar
z(j)}\over\pi}\right)^{n(N-n)}
\nonumber\\
&&\quad
\times\exp\left[{
-\sum_{j=1}^{M}\tr z^{\dagger}(j)\{z(j)
-U(\Delta t)z(j-1)V^{\dagger}(\Delta t)\}
}\right]\ ,
\label{wkba0}
\end{eqnarray}
where
\be
{\tilde{\cal Z}}_{k}(T) \equiv {{\cal Z}_{k}(T) \over (\det V(T))^{k} }  \  .
\label{ztildez}
\ee
To perform the integration, it is convenient to utilize the Fourier
transformation which respects ``PBC'':
\be
z(j)=\sum_{r=0}^{M-1}{1\over\sqrt{M}}e^{-2\pi ijr/M}{\tilde z}(r)\ , \quad
{\tilde z}(r)\in\Mnm{N-n}{n},
\label{fourier1}
\ee
and enables us to write
\bea
&&
\sum_{j=1}^{M} z^{\dagger}(j)\{z(j)
-U(\Delta t)z(j-1)V^{\dagger}(\Delta t)\}
\nonumber\\
&=&
\sum_{r=0}^{M-1} {\tilde z}^{\dagger}(r)\{{\tilde z}(r)
-U(\Delta t){\tilde z}(r)
V^{\dagger}(\Delta t)e^{2\pi ir/M}\}\ .
\label{fourier2}
\eea
Since the Jacobian is trivial the integration with respect to \({\tilde
z}\) can readily be
performed to give
\be
{\tilde{\cal Z}}_{k}(T)
=\lim\limits_{M\rightarrow\infty}
{1\over\prod_{r=0}^{M-1}\det\{1_{N-n}\otimes 1_{n}
-e^{2\pi ir/M}U(\Delta t)\otimes {\bar V}(\Delta t)\}},
\label{fourier3}
\ee
where \({\bar V}\) is the complex conjugate of \(V\).
Recalling the identity which holds for any \(X\in\Mn{m}\),
\be
\prod_{r=0}^{M-1}(1_{m}-e^{2\pi ir/M}X)=1_{m}-X^{M}\ ,
\label{matrixid}
\ee
we finally obtain
\begin{eqnarray}
{\tilde{\cal Z}}_{k}(T)
& = &
{1\over\det\{1_{N-n}\otimes 1_{n}-U(T)\otimes {\bar V}(T)\}}
\nonumber\\
& = &
{1\over{
\prod_{i=n+1}^{N}
\prod_{a=1}^{n}
\{1-e^{-i(h_{i}-h_{a})T}\}
}}\ .
\label{wkba}
\end{eqnarray}

Similar to the classical case there are \({N\choose n}\)'s
classical solutions in total. Therefore taking all the contributions
and going back to the relation \eqref{ztildez}, we obtain
\be
{\cal Z}_{k}(T) \stackrel{ k \rightarrow \infty}{\sim}
\sum_{\mu_{1}<\cdots<\mu_{n}}
{\exp\left({-ik\sum_{a=1}^{n}h_{\mu_{a}}T}\right)
\over
\prod_{a=1}^{n}
\prod_{\nu\in{\bar \mu}}
\{1-e^{-i(h_{\nu}-h_{\mu_{a}})T}\}} \ ,
\label{wkball}
\ee
with \(\bar \mu\) being given by \eqref{barmu}.

If we notice that the right hand side of \eqref{wkball} is rewritten, by means
of
the Laplace  expansion of determinant, as
\be
\text{r.h.s of \eqref{wkball}}
={|\ep^{N-1+k},\ldots,\ep^{N-n+k},\ep^{N-n-1},\ldots,\ep^{1},1|
\over|\ep^{N-1},\ldots,\ep^{1},1|}\ ,
\label{charadet}
\ee
where
\be
|\ep^{m_{1}},\ldots,\ep^{m_{N}}|\equiv
\left\vert
\begin{array}{ccc}
\ep_{1}^{m_{1}}&\cdots&\ep_{1}^{m_{N}}\\
\vdots&\vdots&\vdots\\
\ep_{N}^{m_{1}}&\cdots&\ep_{N}^{m_{N}}
\end{array}
\right\vert\ ,\quad
	\ep_{\mu}\equiv
	e^{-ih_{\mu}T},
\label{defdet}
\ee
we can conclude that the WKB approximation gives an exact result in the case of
the path integral expression \eqref{characf}; since \eqref{charadet} is nothing
but
the Weyl character formula\cite{HW} of \(\Un{N}\) for the present
representation. (It will be evident how to obtain a similar expression
for the classical counter part \eqref{clsadl}.)
 The mechanism of this exactness can be uncovered by means of the
Schwinger boson technique.

\subsection{The mechanism of exactness}
\label{bosepi}
In view of \eqref{characg}, or even after \(M-1\)
integrations as in \eqref{characgb}, the remaining integration looks
still  hard to perform. We know, however,
another recipe for constructing a path integral expression with the aid of the
generalized Schwinger boson.

As was introduced in \eqref{Emunu}, any element \(E_{\mu\nu}\in
u(N)\) has an operator counterpart on \(\cal F\).
Since any \(N\times N\) Hermitian matrix is expandable
in terms of these basis matrices, a quantum Hamiltonian
is given as a self-adjoint operator on \(\cal F\):
\be
{\hat X}\equiv\tr(a^{\dagger}X a)\ ,\
X\in\Hn{N}.
\label{opham}
\ee
Taking the diagonal matrix \(H\) in \eqref{hamiltn} we define
\(\hat{H}\) to interpret the character formula \eqref{characa} as
\be
{\cal Z}_{k}(T)=
\Tr_{k}\exp(-i{\hat H}T)\ ,
\label{charaa}
\ee
in this new formulation. If we use the resolution of unity
\eqref{kanzen3}, we arrive exactly at
the same expression as \eqref{characa}. However, recall that the
operator \(P_{k}\) is not only the  resolution of unity
in the \(k\)-th representation space but also the
projection operator onto the subspace of the Fock space.
Therefore rewrite
\eqref{charaa} as
\bea
{\cal Z}_{k}(T)
& = &
\Tr\{\exp(-i{\hat H}T)
P_{k}\}
\nonumber\\
& = &
\int_{\Un{n}}{dg\over(\det g)^{k}}
\int\left({
dzd{\bar z}\over
\pi
}\right)^{Nn}
\langle Z\vert e^{-i\hat{H}T}\vert Zg_{\veps}\rangle\ ,
\label{charab}
\eea
where we have introduced a regularization parameter \(\veps\), being put
zero
after all:
\be
g\mapsto g_{\veps}=e^{-\veps}g
\label{regul}
\ee
which can legitimatize the exchange of order of integrations. Apart from
the \(g\)-integration, the path integral expression for the right hand
side
of \eqref{charab} can be found straightforwardly: divide
the time
duration into \(M\) segments and insert the resolution of unity
\eqref{bosekanzen} successively to obtain
\bea
&&\int\left({
dzd{\bar z}\over
\pi
}\right)^{Nn}
\langle Z\vert e^{-i\hat{H}T}\vert Zg_{\veps}\rangle
\nonumber\\
&=&
\lim\limits_{M\rightarrow\infty}
\int\limits_{\text{TBC}}\prod_{i=1}^{M}
\left({
dz(i)d{\bar z}(i)\over
\pi
}\right)^{Nn}
\prod_{j=1}^{M}
\langle Z(j)\vert(\bbox{1}-i\hat{H}\Delta t)\vert Z(j-1)\rangle\ ,
\label{gausspatha}
\eea
with ``TBC'' being a twisted boundary condition \(Z(0)=Z(M)g_{\veps}\).
Since the Hamiltonian is bilinear with respect to \(a^{\dagger}\)
and \(a\), it is a simple task to reach
\bea
\text{\eqref{gausspatha}}&=&
\lim\limits_{M\rightarrow\infty}
\int\limits_{\text{TBC}} \prod_{i=1}^{M}
\left({
dz(i)d{\bar z}(i)\over
\pi
}\right)^{Nn}
\nonumber\\
&&\times\exp\left[{
-\sum_{j=1}^{M}\tr Z^{\dagger}(j)\left\{{
Z(j)-(1_{N}-iH\Delta t)Z(j-1)
}\right\}
}\right]\ .
\label{gausspathb}
\eea
In order to carry out the Gaussian path integral in this case, follow a
similar procedure
from \eqref{fourier1} to \eqref{wkba} except employing the Fourier
transformation met
with ``TBC'':
\be
z(j)=\sum_{r=0}^{M-1}{1\over\sqrt{M}}e^{-2\pi ijr/M}{\tilde z}(r) \left(
g_{\veps} \right)^{j
/ M} .
\ee
Therefore
\be
\text{\eqref{charab}}=
\int_{\Un{n}}{dg\over(\det g)^{k}}
{1\over\det(1_{N}\otimes 1_{n}-e^{-iHT}\otimes g_{\veps}^{\cal T})}\ .
\label{charac}
\ee
The remaining \(g\)-integration can be performed by use of the
decomposition\cite{HW2,Mehta}
\be
g=\Omega g_{0} \Omega^{\dagger},\
g_{0}=\text{diag}(e^{i\theta_{1}},\ldots,e^{i\theta_{n}}),\
\Omega\in\SUn{n},
\label{decompg}
\ee
then by the integration with respect to \(\Omega\) giving
\be
\hbox{\rm \eqref{charac}}
=\int_{0}^{2\pi}\left({
d\theta\over2\pi
}\right)^{n}
{1\over n!}
\prod_{a<b}
\vert e^{i\theta_{a}}-e^{i\theta_{b}}\vert^{2}
{\exp\left({-ik\sum_{a=1}^{n}\theta_{a}}\right)
\over
\prod_{\mu=1}^{N}
\prod_{a=1}^{n}
(1-e^{-ih_{\mu}T+i\theta_{a}-\veps})}\ .
\label{charad}
\ee
By putting \(w_{a}=e^{-i\theta_{a}}\), the
integration over the maximal torus is converted into a multiple contour
integrations:
\be
\hbox{\rm \eqref{charad}}
={1\over n!}
\oint\left({
dw\over 2\pi i}\right)^{n}
\prod_{a\ne b}(w_{a}-w_{b})
{\prod_{a=1}^{n}w_{a}^{N-n+k}\over
\prod_{\mu=1}^{N}
\prod_{a=1}^{n}
(w_{a}-e^{-ih_{\mu}T-\veps})}\ .
\label{charaf}
\ee
Taking into account all the contributions from poles, we finally
obtain
\be
{\cal Z}_{k}(T)=
\sum_{\mu_{1}<\cdots<\mu_{n}}
{\exp\left({-ik\sum_{a=1}^{n}h_{\mu_{a}}T}\right)
\over
\prod_{a=1}^{n}
\prod_{\nu\in{\bar \mu}}
\{1-e^{-i(h_{\nu}-h_{\mu_{a}})T}\}}\ ,
\label{generalcase2}
\ee
where again we have used the notation \eqref{barmu}. \eqref{generalcase2}
exactly
matches with \eqref{wkball}. Hence the WKB approximation is exact in path
integral for the
character formula \eqref{characa} which is now interpreted as that of
\(\Un{N}\)
represented over \(\grsmn{N}{n}\).

Here, the reason is quite
obvious; since the path
integral representation, in view of \eqref{gausspathb}, is essentially
Gaussian with  an additional \(g\)-integration which is regarded as
imposing the physical
state condition.  Evidently there is no room for the appearance of
\(k^{-1}\). As
is stated in \cite{BT}, we may conclude that the path integral expression
we have
discussed is
kinematically nonlinear but dynamically free. (The situation would
correspond to
a free field over nontrivial phase space, which should compare to the
harmonic oscillator (free field!) over a flat phase space.)

\section{Discussion}
\label{discuss}

In this paper we have clarified the exactness of the WKB approximation
for the \(\Un{N}\) character formula which is formulated by
path integral over \(\grsmn{N}{n}\). We have
employed a time slicing method and coherent states to build up a  path
integral representation. We have made two
different approaches:  Perelomov's generalized coherent state and a
generalized method of
Schwinger boson. In terms of the latter method, that is, a view from a
constrained system
clarifies the reason for exactness: both cases, classical \eqref{clsadl}
as well as
\eqref{wkball}, can be interpreted such that the targets, the classical
partition function
\eqref{partita} and the character formula \eqref{characa}, have
essentially been expressed as Gaussian forms.

Note, however, the difference: while critical points are
controlled by eigenvalues of the Hermitian matrix,
\eqref{eigen1} or \eqref{clsadl} in the
classical case, but those are controlled by those of the unitary matrix,
\eqref{eqofmotiona}, \eqref{eqofmotionb} or \eqref{wkball}
in the quantum case.
This originates from the difference in the form of constrains:
in the classical case we can naively
put constraints
\eqref{constra} into a trivial partition function in terms of the delta
function to obtain
\eqref{partic}, in other words, the integration domain of the multiplier
\(\lambda\) is infinite
but in the quantum case, as can be seen from
\eqref{projection}, it is {\em compact}.

The compactness of the integration domain is indispensable
in the quantum case: to see the situation more clearly let us examine the
following model. Take the \(\bbox{C}P^{1}\) case as a simple example:
\be
H=\boldkey{z}^{\dagger}h\boldkey{z}\ ,\quad
h=
\pmatrix{
a&b\cr
\bar{b}&d
}\in\Hn{2}\ ,\quad
\boldkey{z}=
\pmatrix{z_{1}\cr
z_{2}}\in\boldkey{C}^{2}\ ,
\label{cp1ham}
\ee
with a constraint
\be
\psi\equiv\boldkey{z}^{\dagger}\boldkey{z}-p\approx 0,\quad
p\in\boldkey{R}_{+}.
\label{cp1const}
\ee
The fundamental Poisson brackets are given by
\be
\{z_{\mu},\bar{z}_{\nu}\}=-i\delta_{\mu\nu},\
\{z_{\mu},z_{\nu}\}=\{\bar{z}_{\mu},\bar{z}_{\nu}\}=0,\quad
(\mu,\nu=1,2)\ .
\label{cp1pbra}
\ee
In order to reach the reduced manifold \(\bbox{C}P^{1}\) we need an additional
constraint to fix one phase of complex numbers, say \(z_1\). To this end, a
change of variables
\be
\boldkey{z}=\pmatrix{
1\cr
\xi}{1\over\sqrt{1+|\xi|^{2}}}\zeta\ ,
\label{chvar}
\ee
is utilized. The constraint \eqref{cp1const} is read as
\(\psi=|\zeta|^{2}-p\) and the
desired one is found as
\be
\chi={1\over 2i}\log(\zeta/\bar{\zeta})-\phi_{0}, \quad
0\le\phi_{0}<2\pi \ .
\label{chicond}
\ee
They satisfy
\be
\{\psi,\chi\}=1\ ,
\label{cnspbra}
\ee
so that the Dirac brackets can be constructed giving
\be
\{\xi,\bar{\xi}\}_{\rm D}=-{i\over p}(1+|\xi|^{2})^{2}.
\label{cp1dirac}
\ee
In this way classical mechanics on the reduced phase space can be obtained
without
any problems.

A prescription to quantum theory would be found by means of path integral
developed by Faddeev-Senjanovic(FS)\cite{FS}:
\bea
{\cal Z}^{(\rm FS)}_p
& \equiv &
\lim\limits_{M\rightarrow\infty}
\int\limits_{\text{PBC}}
\prod_{i=1}^{M}{(dz(i)d\bar{z}(i))^{2}\over\pi}
\delta(\psi(i))\delta(\chi(i))
\nonumber\\
&&
\times\exp\left[{i\sum_{j=1}^{M}\left\{{
i\boldkey{z}^{\dagger}(j)\Delta\boldkey{z}(j)
-\Delta t\boldkey{z}^{\dagger}(j)h\boldkey{z}(j-1)
}\right\}
}\right] \ ,  \label{FSformula}
\eea
where
\be
\Delta\boldkey{z}(j) \equiv \boldkey{z}(j)-\boldkey{z}(j-1)  \ .
\ee
Although the way of finding \eqref{FSformula} is rather heuristic, the
result seems
convincing provided constrained systems are given in the configuration
space such as
the sphere\cite{FK2}. Therefore we employ this as a starting point of
quantum theory.

With the aid of the change of variables \eqref{chvar}, \eqref{FSformula}
becomes
\bea
{\cal Z}^{(\rm FS)}_{p}
& = &
\lim\limits_{M\rightarrow\infty}
\int\limits_{\text{PBC}}
\prod_{i=1}^{M}{d\xi(i)d\bar{\xi}(i)\over\pi(1+|\xi(i)|^{2})^{2}}
d\zeta(i)d\bar{\zeta}(i)|\zeta(i)|^{2}
\delta(\psi(i))\delta(\chi(i))
\nonumber\\
&&\!\!\!\!\!\!\!\!\!\!\!\!\!\!\!\!\!\!\!\!\!
\times\exp
\left[{  i \sum_{j=1}^{M} \left\{ {
i |\zeta(j)|^{2}
-i\bar{\zeta}(j)\zeta(j-1){
1+\bar\xi(j)\xi(j-1)\over
(1+\bar\xi(j)\xi(j))^{1/2}(1+\bar\xi(j-1)\xi(j-1))^{1/2}} }\right. }\right.
\nonumber \\
&&
\left.{\left.{
-\Delta t\bar{\zeta}(j)\zeta(j-1)
{a+b\xi(j-1)+\bar{\xi}(j)\bar{b}+\bar{\xi}(j)d\xi(j-1)\over
(1+\bar\xi(j)\xi(j))^{1/2}(1+\bar\xi(j-1)\xi(j-1))^{1/2}} }\right\}
}\right]   \ .
\label{cp1ptf}
\eea
A trivial integration with respect to \(\zeta\) leads to
\bea
{\cal Z}^{(\rm FS)}_{p}
& = &
\lim\limits_{M\rightarrow\infty}
\int\limits_{\text{PBC}}
\prod_{i=1}^{M}{pd\xi(i)d\bar{\xi}(i)\over\pi(1+|\xi(i)|^{2})^{2}}
\nonumber\\
&&\!\!\!\!\!\!\!\!\!\!\!\!\!\!\!\!\!\!\!\!\!
\times\exp
\left[{  -p \sum_{j=1}^{M} \left\{ {
1-{
1+\bar\xi(j)\xi(j-1)\over
(1+\bar\xi(j)\xi(j))^{1/2}(1+\bar\xi(j-1)\xi(j-1))^{1/2}} }\right. }\right.
\nonumber \\
&&
\left.{\left.{
+i\Delta t
{a+b\xi(j-1)+\bar{\xi}(j)\bar{b}+\bar{\xi}(j)d\xi(j-1)\over
(1+\bar\xi(j)\xi(j))^{1/2}(1+\bar\xi(j-1)\xi(j-1))^{1/2}} }\right\}
}\right]   \  ,
\label{cp1ptf2}
\eea
which should be compared with the correct one, \(\bbox{C}P^{1}\) version of
\eqref{characf},
\bea
{\cal Z}_{k}(T)
& = &
\lim\limits_{M\rightarrow\infty}
\int\limits_{\text{PBC}} \prod_{i=1}^{M}
{(k+1)d\xi(i)d\bar{\xi}(i)\over\pi(1+|\xi(i)|^{2})^{2}}
\nonumber\\
&&
\times
\exp\left[{
-k\sum_{j=1}^{M}
\left\{{
\log(1+|\xi(j)|^{2})
-\log(1+\bar{\xi}(j)\xi(j-1))
\vphantom{
{a+b\xi(j-1)+\bar{\xi}(j)\bar{b}+\bar{\xi}(j)d\xi(j-1)\over
1+\bar{\xi}(j)\xi(j-1)}
}
}\right.
}\right.
\nonumber\\
& &
\left.{\left.{
\hphantom{
\times
\exp
}
+ i\Delta t
{a+b\xi(j-1)+\bar{\xi}(j)\bar{b}+\bar{\xi}(j)d\xi(j-1)\over
1+\bar{\xi}(j)\xi(j-1)}
}\right\}
}\right]\ .
\label{cp1characf}
\eea
In view of these, even if an arbitrary parameter \(p\) in \eqref{cp1ptf2}
would be set to \(k \in \bbox{Z}_{+}\), a failure of the FS prescription
for the present model is  now obvious.
Nevertheless a formal continuum limit of
\eqref{cp1ptf2} seems reasonably geometric and respect the classical
feature of the system. First rely on a naive expansion,
\be
\xi(j-1)  \sim \xi(j)  - \Delta t {\dot{\xi}(j) } \ ,
 \label{cp1naive}
\ee
which brings \eqref{cp1ptf2} to
\bea
\text{\eqref{cp1ptf2}}
& \rightarrow &
\int\limits_{\text{PBC}}
\prod_{0\le t\le T}{pd\xi(t)d\bar{\xi}(t)\over\pi(1+|\xi(t)|^{2})^{2}}
\nonumber\\
&&
\!\!\!\!\!\!\!\!\!\!\!\!\!\!\!\!\!\!
\times\exp\left[{
ip\int_{0}^{T}dt\left\{{
{i\over2}{\bar{\xi}\dot{\xi}-\dot{\bar{\xi}}\xi\over 1+|\xi|^{2}}
-{1\over 1+|\xi|^{2}}(a+b\xi+ \bar{b} \bar{\xi}+ \bar{\xi}d\xi)
}\right\}
}\right]\ ,
\label{cp1cnt}
\eea
whose exponent consists of (classical) \(\bbox{C}P^{1}\) action. (Of
course, we can arrive
also at \eqref{cp1cnt}, starting from \eqref{cp1characf} with an replacement of
\(k+1\) in the measure by
\(p\) and taking the same limit.)

We should, therefore, discard or modify the expression \eqref{FSformula}
in order to
find a correct quantum theory.  As was stated above, to see the importance
of
the compactness of the multiplier we employ a modified expression:
\bea
{\cal Z}_{p}^{(\text{FS-I})}
& \equiv &
\lim\limits_{M\rightarrow\infty}
\int\limits_{\text{PBC}}
\prod_{i=1}^{M}\left({dz(i)d\bar{z}(i)\over\pi}\right)^{2}
\!\!\!\int_{-\infty}^{\infty}
{\Delta t\over2\pi}d\lambda(i)
\exp\left[{i\sum_{j=1}^{M}\left\{{
i\boldkey{z}^{\dagger}(j)\Delta \boldkey{z}(j)
}\right.}\right.
\nonumber\\
&&
\left.{
\hphantom{\lim\limits_{M\rightarrow\infty}}
\vphantom{\sum_{j=1}^{M}}\left.{
-\Delta t\boldkey{z}^{\dagger}(j)h\boldkey{z}(j-1)
+\Delta t\lambda(j)(\boldkey{z}^{\dagger}(j)\boldkey{z}(j-1)-p)
}\right\}
}\right]\ .
\label{cp1delt}
\eea
Here the \(\chi\) constraints have simply been discarded while the \(\psi\)
constraints now read as
\(\boldkey{z}^{\dagger}(j)\boldkey{z}(j-1)-p \approx 0\)
and have been Fourier-transformed in \eqref{FSformula}.
Note that
\(\lambda\) plays a role of the multiplier and still travels an
{\em infinite range\/}. If we  notice that a change of variables
\bea
\boldkey{z}(j)&=&\boldkey{z}^{\prime}(j)
\exp\left\{i\Delta t\sum_{k=1}^{j}\lambda(k)\right\}\ ,
\label{cp1chvar1}\\
\lambda(j)&=&\sum_{r=0}^{M-1}{1\over\sqrt{M}}e^{-2\pi
ijr/M}\tilde{\lambda}(r)
\ ,  \label{cp1chvar2}
\eea
wipes out almost all \(\tilde{\lambda}(r)\) leaving only
\(\tilde{\lambda}(0)\)(constant mode of \(\lambda(j)\)) in the
integrand we further modify \eqref{FSformula}, by throwing away
infinities from \(\tilde{\lambda}(r)\)'s, to
\bea
{\cal Z}_{p}^{(\text{FS-II})}
& \equiv &
\lim\limits_{M\rightarrow\infty}
\int\limits_{\text{TBC}}
\prod_{i=1}^{M}\left({dz(i)d\bar{z}(i)\over\pi}\right)^{2}
\int_{-\infty}^{\infty}
{d\lambda\over2\pi}
e^{-ip\lambda}
\nonumber\\
&&
\times
\exp\left[{i\sum_{j=1}^{M}\left\{{
i\boldkey{z}^{\dagger}(j)\Delta \boldkey{z}(j)
-\Delta t\boldkey{z}^{\dagger}(j)h\boldkey{z}(j-1)
}\right\}
}\right]\ ,
\label{cp1atotyotto}
\eea
where as before ``TBC'' denotes
\(\boldkey{z}(0)=\boldkey{z}(M)e^{i\lambda}\)
and all primes have been removed.
Now the Gaussian integrations with respect to
\(\boldkey{z}\)'s can be done,  by introducing a regularization parameter
\(\veps>0\), to
yield
\be
{\cal Z}_{p}^{(\text{FS-II})}
=\lim\limits_{\veps\rightarrow\infty}
\int_{-\infty}^{+\infty}{d\lambda\over2\pi}e^{-ip\lambda}
{1\over(1-e^{-ih_{1}T+i\lambda-\veps})(1-e^{-ih_{2}T+i\lambda-\veps})}\ ,
\label{cp1saigo}
\ee
where \(h_{i}\)'s are eigenvalues of \(h\) in \eqref{cp1ham}.
In view of  \eqref{cp1saigo}, \(p\) must be some positive integer,
otherwise the result is zero, which leads us furthermore to the
conclusion that the integration domain of \(\lambda\) in
\eqref{cp1saigo} {\em must be replaced by a compact one} \(0\le \lambda
\le 2\pi\).
(Since otherwise we obtain infinite copies of the same integration.)
Therefore we should have
\be
{\cal Z}_{p}^{(\text{correct})} = \lim\limits_{\veps\rightarrow\infty}
\int_{0}^{2\pi}{d\lambda\over2\pi}e^{-ip\lambda}
{1\over(1-e^{-ih_{1}T+i\lambda-\veps})(1-e^{-ih_{2}T+i\lambda-\veps})} \ ,
\label{cp1correcta}
\ee
that is
\bea
{\cal Z}_{p}^{(\text{correct})}
& = &
\lim\limits_{M\rightarrow\infty}
\int\limits_{\text{TBC}}
\prod_{i=1}^{M}\left({dz(i)d\bar{z}(i)\over\pi}\right)^{2}
\int_{0}^{2\pi}
{d\lambda\over2\pi}
e^{-ip\lambda}
\nonumber\\
&&
\times
\exp\left[{i\sum_{j=1}^{M}\left\{{
i\boldkey{z}^{\dagger}(j)\Delta \boldkey{z}(j)
-\Delta t\boldkey{z}^{\dagger}(j)h\boldkey{z}(j-1)
}\right\}
}\right] \ .
\label{cp1correctb}
\eea
In this way the importance of the compactness in the domain of multipliers
can be
recognized, which convinces us that the use of projection operator
\(P_{k}\)
\eqref{projection} given in section \ref{scwcstrctn} is indispensable.

\appendix
\section{Proof of the theorems}
\label{appa}
In this appendix, we prove our main theorems in section \ref{algcstrctn}.

\subsection{Theorem \protect\ref{excellent1}}
\label{appa1}
The statement is
\be
\int_{\Un{n}}{dg\over (\det g)^{p}}\exp\{\tr (gX)\}
={\cal N}(n,p)|X|^{p},
\label{appexcellenta}
\ee
with
\be
{\cal N}(n,p)={{0!1!\cdots(n-1)!}\over {p!(p+1)!\cdots(p+n-1)!}}\ ,
\label{appnnp}
\ee
and the assumptions being given in the text.

We use the following facts without proof.
\begin{enumerate}
\renewcommand{\labelenumi}{\itembox{\Roman{enumi}}}

\item Invariant measure on \(\Un{n}\):
\be
dg\propto {1\over{(\det g)}^{n}}\prod_{1\le i,j\le n}dg_{ij}\ ,
\label{unmeasure}
\ee
where \(dg_{ij}\)'s denote \(n^{2}\) independent differentials.
Each \(g_{ij}\) is complex  and the number of independent components is
\(n^2\) in terms of
real variables.
\item Local decomposition of \(\Un{n}\):
\be
g\in \Un{n}\Rightarrow
g=\left(
\begin{array}{cc}
a&0\\
0&B
\end{array}
\right)
\exp
\left(
\begin{array}{cc}
0&-\alpha^{\dagger}\\
\alpha&0
\end{array}
\right)\ ,
\label{unbunkai}
\ee
where \(a\in \Un{1},\ B\in \Un{n-1}\) and \(\alpha\in \bbox{C}^{n-1}\)
 is the parameter for \(\bbox{C}P^{n-1}\). Rewrite \eqref{unbunkai} to
\be
\left(
\begin{array}{cc}
a&0\\
0&B
\end{array}
\right)
\left(
\begin{array}{cc}
\dfrac{1}{\sqrt{1+|\xi|^{2}}}&-\dfrac{1}{\sqrt{1+|\xi|^{2}}}\xi^{\dagger}\\
\xi\dfrac{1}{\sqrt{1+|\xi|^{2}}}&\dfrac{1}{\sqrt{1_{n-1}+\xi\xi^{\dagger}}}
\end{array}
\right),\quad
\xi={\alpha\over|\alpha|}\tan|\alpha|\ ,
\label{unbunkai2}
\ee
so that
\be
dg\propto d\mu_{n-1}(\xi){da\over a}{\prod dB_{ij}\over(\det
B)^{n-1}}\ ,
\label{measureun}
\ee
where the measure has been decomposed into \(\bbox{C}P^{n-1},\ \Un{1}\)
and
\(\Un{n-1}\) in that order. Therefore a repeated use of the procedure
results in
\be
dg\propto \prod_{j=1}^{n-1}d\mu_{j}(\xi^{(j)})\prod_{i=1}^{n}{da_{i}\over
a_{i}}\ ,
\ee
that is, the invariant measure of \(\Un{n}\) is given by the product of
 \(\bbox{C}P^{j}\)'s measure(\( 1\le j\le n-1\)) and the tori of \(\Un{n}\),
which corresponds to the local decomposition of \(\Un{n}\):
\bea
&&{\Un{n}\over{\Un{1}\times\Un{n-1}}}\times
{\Un{n-1}\over{\Un{1}\times\Un{n-2}}}
\times\cdots\times
{\Un{2}\over{\Un{1}\times\Un{1}}}
\times {\Un{1}}^{n}
\nonumber\\
&\cong&
\bbox{C}P^{n-1}\times\bbox{C}P^{n-2}\times\cdots
\bbox{C}P^{1}\times {\Un{1}}^{n}\ .
\label{unbunkailocal}
\eea

\item Integration formula on \(\bbox{C}P^{N}\):
\be
{(k+N)!\over k!}\int{(d\xi d{\bar
\xi})^{N}\over\pi^{N}(1+|\xi|^{2})^{N+1+k}}
(1+a^{\dagger}\xi)^{k}(1+\xi^{\dagger} b)^{k}
=(1+a^{\dagger}b)^{k}
\label{cpnfrm}
\ee
holds for \({}^{\forall} a,b\in\bbox{C}^{N}\) and
\({}^{\forall}k\in\bbox{Z}_{+}\).
\end{enumerate}

Since both sides of \eqref{appexcellenta} are regular functions of
\(x_{ij}\), the case, \(|X|=0\), can be regarded as a limit of \(|X|\ne
0\).
Then \(X\) can be assumed without loss of generality as
\be
X=\left(\begin{array}{cc}
\alpha&\beta\\
\gamma&\delta
\end{array}\right),\quad
\alpha\in \bbox{C}, \beta^{T},\gamma\in \bbox{C}^{n-1},\delta\in
M(n-1;\bbox{C})\ ,
\ee
with \(\alpha\det \delta\ne 0\) .

The proof is done by induction:
\begin{enumerate}
\renewcommand{\labelenumi}{\itembra{\Roman{enumi}}}
\item For \(n=1\), \eqref{appexcellenta} is verified by a direct
calculation:
\be
\oint{da\over 2\pi ia^{k+1}}
\sum_{k=0}^{\infty}
{1\over k!}(aX)^{k}
={X^{k}\over k!}\ .
\ee
\item Assume \eqref{appexcellenta} holds for \(n\le m\). Then adopt
\eqref{unbunkai2} for \(g\in\Un{m+1}\) to find
\be
\tr g\left(\begin{array}{cc}
\alpha&\beta\\
\gamma&\delta
\end{array}\right)
=a{\alpha-\xi^{\dagger}\gamma\over\sqrt{1+|\xi|^{2}}}
+\tr\left\{{
B\left({
{\xi \beta \over\sqrt{1+|\xi|^{2}}}+
{1\over\sqrt{1_{m}+\xi\xi^{\dagger}}}\delta
}\right)
}\right\}\ .
\ee
According to the assumption of induction, when \(n=m+1\), the integration
with respect
to \(a\) and \(B\) in  the left hand side of \eqref{appexcellenta} gives
\bea
&&{{\cal N}(m,p)\over p!}\int{m!(d\xi d{\bar \xi})^{m}
\over\pi^{m}(1+|\xi|^{2})^{m+1}}
\left\{
{\alpha-\xi^{\dagger}\gamma\over\sqrt{1+|\xi|^{2}}}
\right\}^{p}
\nonumber\\
&&\hphantom{{{\cal N}(m,p)\over p!}}
\times
\left\{{
\det\left({
{\xi \beta \over\sqrt{1+|\xi|^{2}}}+
{1\over\sqrt{1_{m}+\xi\xi^{\dagger}}}\delta
}\right)
}\right\}^{p}\ .
\label{step1}
\eea
By means of a relation
\be
\det\left({
{\xi \beta \over\sqrt{1+|\xi|^{2}}}+
{1\over\sqrt{1_{m}+\xi\xi^{\dagger}}}\delta
}\right)
=
{1+\beta\delta^{-1}\xi\over\sqrt{1+|\xi|^{2}}}\det \delta\ ,
\label{junbi}
\ee
\eqref{step1} is rewritten as
\be
{m!{\cal N}(m,p)\over p!}
(\alpha\det\delta)^{p}
\int{(d\xi d{\bar \xi})^{m}
\over\pi^{m}(1+|\xi|^{2})^{m+1+p}}
(1-\xi^{\dagger}\gamma\alpha^{-1})^{p}(1+\beta\delta^{-1}\xi)^{p} \ ,
\label{step2}
\ee
which finally turns out, with the aid of \eqref{cpnfrm}, to be
\begin{eqnarray}
\text{\eqref{step2}}&=&
{m!\over(p+m)!}{\cal N}(m,p)
(\alpha\det \delta)^{p}(1-\beta\delta^{-1}\gamma\alpha^{-1})^{p}
\nonumber\\
&=&
{m!\over(p+m)!}{\cal N}(m,p)
\left\{{
\det
\left(\begin{array}{cc}
\alpha&\beta\\
\gamma&\delta
\end{array}\right)
}\right\}^{p}\ .
\end{eqnarray}
Hence \eqref{appexcellenta} holds for \(n=m+1\) as well as for \(n\le m\).
\end{enumerate}
This completes the proof.

\subsection{Theorem \protect\ref{excellent2}}
\label{appa2}
Next consider the second theorem:
\begin{eqnarray}
\left\vert{\partial_{X}}\right\vert
\left\vert X\right\vert^{p}
& = &
p(p+1)\cdots(p+n-1)
\left\vert X\right\vert^{p-1} \ ,
\nonumber\\
p&=&0,\pm 1,\pm 2,\ldots  \  .
\label{appformulaa}
\end{eqnarray}

\subsubsection{Proof for \(p\ge 0\).}
In this case, the formula \eqref{appexcellenta} is utilized to rewrite the
left hand side of
\eqref{appformulaa} as
\begin{equation}
\left\vert{\partial_{X}}\right\vert
\left\vert X\right\vert^{p}
=
{1\over{\cal N}(n,1){\cal N}(n,p)}
\int_{\Un{n}}{dg_{1}dg_{2}\over \det g_{1}(\det g_{2})^{p}}
\exp\left\{{
\tr (g_{1}\partial_{X})
}\right\}
\exp\left\{{
\tr (g_{2}X)
}\right\}.
\label{integralrep}
\end{equation}
Regard \(\partial_{ij}=\partial/\partial x_{ij}\) and \(x_{ij}\) as
operators upon functions of \(x_{ij}\) so that the both sides of
\eqref{integralrep} are implied as acting on \(1\). Then use is made of the
Campbell-Baker-Hausdorff formula in the right hand side to interchange two
exponential
factors:
\bea
\text{r.h.s. of \eqref{integralrep}}
& = &
{1\over{\cal N}(n,1){\cal N}(n,p)}
\int_{\Un{n}}{dg_{1}dg_{2}\over \det g_{1}(\det g_{2})^{p}}
\nonumber \\
& &
\times
\exp\left\{{
\tr (g_{1}g_{2})
}\right\}
\exp\left\{{
\tr (g_{2}X)
}\right\}
\exp\left\{{
\tr (g_{1}\partial_{X})
}\right\}\ ,
\label{intrepb}
\eea
so that the last exponential factor can be dropped. Finally \(g_{1}\)
integration leads to
\begin{eqnarray}
\left\vert{\partial_{X}}\right\vert
\left\vert X\right\vert^{p}
& = &
{1\over{\cal N}(n,p)}
\int_{\Un{n}}{dg\over (\det g)^{p-1}}
\exp\left\{{
\tr (gX)
}\right\}
\nonumber\\
& = &
{{\cal N}(n,p-1)\over{\cal N}(n,p)}|X|^{p-1}
\nonumber\\
& = &
p(p+1)\cdots(p+n-1)|X|^{p-1},
\label{profpos}
\end{eqnarray}
which complete the proof for \(p\in\bbox{Z}_{+}\). Note that there is no
restriction to
\(X\in\Mn{n}\) in this case.

\subsubsection{Proof for \(p< 0\).}
In this case, note the following relation:
\begin{eqnarray}
\left\vert{\partial_{X}}\right\vert
\left\vert X\right\vert^{-p}
& = &
\left\vert{\partial_{X}}\right\vert \int\left({dzd\bar
z\over\pi}\right)^{np}
\exp\left\{-\tr (XZZ^{\dagger})\right\}
\nonumber\\
& = &
(-1)^{n}
\int\left({dzd\bar z\over\pi}\right)^{np}
\vert{ZZ^{\dagger}}\vert
\exp\left\{-\tr (XZZ^{\dagger})\right\}\ ,
\label{negab}
\end{eqnarray}
for
\be
X=A+iB, \quad A^{\dagger}=A, \quad  B^{\dagger}=B, \quad A>0 \ ,
\label{joken}
\ee
where we have put \(p\rightarrow -p\) so that \(p\) is positive here and
hereafter.

First consider the case \(p\le n-1\). If we notice a relation
\be
\det ZZ^{\dagger}=
\det
\pmatrix{
Z&0
}
\pmatrix{
Z^{\dagger}\cr
0}
=0\ ,
\label{plessthann}
\ee
we find that the right hand side of \eqref{negab} trivially vanishes when
\(p=0,\ldots,n-1\).
Hence it is enough to examine the case \(p\ge n\).  Before proceeding, we
recall a well
known fact: under  the condition \eqref{joken} \(A\) and
\(B\) are simultaneously diagonalized by means of an appropriate
invertible matrix \(K\)
such that
\be
K^{\dagger}AK=1_{n},\quad
K^{\dagger}BK=B_{D}=\text{diag}(\b_{1},\ldots,\b_{n})\ .
\ee
Accordingly a change of variables, \(Z\mapsto
Z^{\prime}=K^{-1}Z\), gives
\be
\text{\eqref{negab}}={(-1)^{n}\over |A|^{p+1}}
\int\left({dzd\bar z\over\pi}\right)^{np}
\vert{ZZ^{\dagger}}\vert
\exp\left[-\tr \{(1_{n}+iB_{D})ZZ^{\dagger}\}\right].
\label{negabb}
\ee
Then rewrite the matrix \(1_{n}+iB_{D}\) as
\begin{equation}
1_{n}+iB_{D}=
\left(\begin{array}{ccc}
1+i\b_{1}&&0\\
&\ddots&\\
0&&1+i\b_{n}
\end{array}
\right)
=F\Phi,
\label{scalinga}
\end{equation}
\begin{equation}
F\equiv
\left(\begin{array}{ccc}
f_{1}&&0\\
&\ddots&\\
0&&f_{n}
\end{array}
\right),\quad
\Phi\equiv
\left(\begin{array}{ccc}
e^{i\phi_{1}}&&0\\
&\ddots&\\
0&&e^{i\phi_{n}}
\end{array}
\right),
\label{scalingb}
\end{equation}
\be
0< f_{i},\qquad  -{\pi\over 2}<\phi_{i}<{\pi\over 2} \quad
(i=1,\ldots,n)\ .
\ee
Further a change of variables \(Z\mapsto Z^{\prime}=\sqrt{F}Z\) leads to
\be
\hbox{\rm \eqref{negabb}}=
{(-1)^{n}\over |A|^{p+1}|F|^{p+1}}
\int\left({dzd\bar z\over\pi}\right)^{np}
\vert{ZZ^{\dagger}}\vert
\exp\left\{-\tr (\Phi ZZ^{\dagger})\right\}\ ,
\label{negabc}
\ee
which is rewritten, by use of the formula \eqref{appexcellenta}, as
\be
\hbox{\rm \eqref{negabc}}=
{(-1)^{n}n!\over |A|^{p+1}|F|^{p+1}}
\lim\limits_{\veps\rightarrow 0}
\int_{\Un{n}}{dg\over \det g}
\int\left({dzd\bar z\over\pi}\right)^{np}
\exp\left[-\tr \{(\Phi-ge^{-\veps}) ZZ^{\dagger}\}\right]\ .
\label{negabd}
\ee
Now the Gaussian integration with respect to \(Z\) is performed to be
\be
\hbox{\rm \eqref{negabd}}={(-1)^{n}n!\over |A|^{p+1}|F|^{p+1}}
\lim\limits_{\veps\rightarrow 0}
\int_{\Un{n}}{dg\over \det g}{1\over{\det(\Phi-ge^{-\veps})^{p}}}\ ,
\label{negabda}
\ee
which becomes after a change of variable \(g\mapsto \Phi^{-1} g\) to
\begin{eqnarray}
\hbox{\rm \eqref{negabda}}& = &
{(-1)^{n}n!\over |A|^{p+1}|F|^{p+1}|\Phi|^{p+1}}
\lim\limits_{\veps\rightarrow 0}
\int_{\Un{n}}{dg\over \det g}{1\over{\det(1_{n}-ge^{-\veps})^{p}}}
\nonumber\\
& = &
{(-1)^{n}n!\over |X|^{p+1}}
\lim\limits_{\veps\rightarrow 0}
\int_{\Un{n}}{dg\over \det g}{1\over{\det(1_{n}-ge^{-\veps})^{p}}}\ .
\label{negabe}
\end{eqnarray}
In view of \eqref{negabe} and \eqref{appformulaa} (with \(p \rightarrow
-p\)), our
remaining task is therefore to prove
\be
\lim\limits_{\veps\rightarrow 0}
\int_{\Un{n}}{dg\over \det g}{1\over{\det(1_{n}-ge^{-\veps})^{p}}}
={p\choose n}\ .
\ee

In order to perform the \(g\)-integration, recall the
decomposition\cite{HW2,Mehta},
\be
g=\Omega g_{0} \Omega^{\dagger},\
g_{0}=\text{diag}(e^{i\theta_{1}},\ldots,e^{i\theta_{n}}),\
\Omega\in\SUn{n},
\label{appdecompg}
\ee
and integrate \(\Omega\) to obtain
\begin{eqnarray}
&&\lim\limits_{\veps\rightarrow 0}
\int_{\Un{n}}{dg\over \det g}{1\over{\det(1_{n}-ge^{-\veps})^{p}}}
\nonumber\\
&=&
\lim\limits_{\veps\rightarrow 0}
\int_{0}^{2\pi}\left({d\theta\over2\pi}\right)^{n}
{1\over n!}\prod_{a<b}|e^{i\theta_{a}}-e^{i\theta_{b}}|^{2}
\prod_{a=1}^{n}{e^{-i\theta_{a}}\over(1-e^{-\veps+i\theta_{a}})^{p}}
\nonumber\\
&=&
\lim\limits_{\veps\rightarrow 0}
\int_{0}^{2\pi}\left({d\theta\over2\pi}\right)^{n}
{1\over n!}\sum_{\displaystyle \sigma,\tau\in{\cal S}_{n}}
\sgn{\sigma\tau}
\prod_{a=1}^{n}\sum_{l_{a}=0}^{\infty}
{p+l_{a}-1\choose l_{a}}
\nonumber\\
&&
\hphantom{\lim\limits_{\veps\rightarrow 0}
\int_{0}^{2\pi}\left({d\theta\over2\pi}\right)^{n}
{1\over n!}}
\times
\exp\left[{
i\left\{{
l_{a}-1+\sigma(a)-\tau(a)
}\right\}\theta_{a}-\l_{a}\veps
}\right]\ ,
\label{inttorus}
\end{eqnarray}
which then yields after the \(\theta\)'s integrations and putting
\(\veps\rightarrow 0\) to
\be
\text{\eqref{inttorus}}=
\left\vert
\begin{array}{ccccc}
p&\dchoose{p+1}{2}&\dchoose{p+2}{3}&\cdots&\dchoose{p+n-1}{n}\\
1&p&\dchoose{p+1}{2}&\ddots&\vdots\\
0&1&p&\ddots&\dchoose{p+2}{3}\\
\vdots&\ddots&\ddots&\ddots&\dchoose{p+1}{2}\\
0&\cdots&0&1&p
\end{array}
\right\vert\ .
\label{negaf}
\ee
We denote this determinant by \({\cal D}(n,p),\ n=0,1,2,\ldots\)
with defining \({\cal D}(0,p)=1\). Recall that our goal is now to show
\begin{equation}
{\cal D}(n,p)={p\choose n}
\label{negah}
\end{equation}
Let us prove this again by induction: first notice the recursion relation,
obtained by an
expansion in the first row of \eqref{negaf},
\begin{equation}
{\cal D}(n,p)=\sum_{r=1}^{n}(-1)^{1+r}{p+r-1\choose r}{\cal D}(n-r,p)\ .
\label{negag}
\end{equation}
Assume \eqref{negah} for \(0\le m\le n-1\),
then \eqref{negag} reads
\begin{eqnarray}
{\cal D}(n,p)
& = &
\sum_{r=1}^{n}(-1)^{1+r}{p+r-1\choose r}{p\choose n-r}
\nonumber\\
& = &
{p\over n}
\sum_{r=1}^{n}(-1)^{1+r}{n\choose r}{p+r-1\choose p+r-n}\ .
\label{negai}
\end{eqnarray}
Utilizing the generating function
\begin{equation}
{p+r-1\choose p+r-n}
=
{1\over p!}\left.{\left({d\over dx}\right)^{p}}\right\vert_{x=0}
\sum_{l=0}^{\infty}{l+n-1\choose l}x^{l+n-r}\ ,
\label{negaj}
\end{equation}
we find
\begin{eqnarray}
&&{p\choose n}-{\cal D}(n,p)
={p\over n}\sum_{r=0}^{n}(-1)^{r}
{n\choose r}{p+r-1\choose p+r-n}
\nonumber\\
&=&
{p\over n}{1\over p!}
\left.{\left({d\over dx}\right)^{p}}\right\vert_{x=0}
\sum_{r=0}^{n}(-1)^{r}{n\choose r}\sum_{l=0}^{\infty}{l+n-1\choose
l}x^{l+n-r}
\nonumber\\
& = &
{p\over n}{1\over p!}
\left.{\left({d\over dx}\right)^{p}}\right\vert_{x=0}
\sum_{r=0}^{n}{n\choose r}(-1)^{r}x^{n-r}
\left({1\over 1-x}\right)^{n}
\nonumber\\
& = &
(-1)^{n}{p\over n}{1\over p!}
\left.{\left({d\over dx}\right)^{p}}\right\vert_{x=0}
1
\nonumber\\
& = &
0\ .
\end{eqnarray}
Thus we have found that \eqref{negah} is also valid for \(m=n\). This
completes the proof.

\section{Feynman kernel and the WKB approximation}
\label{Feynman}
In order to make a clear connection to the D-H theorem, we have
concentrated on the character formula in the text. However, from the
quantum
mechanical point of view, the Feynman kernel is regarded primitive so
that in this
appendix a brief sketch is presented to show the way to a path integral
representation and discuss the WKB approximation.

\subsection{Derivation of the Feynman kernel (method 1)}
Take a Hermitian matrix,
\be
H\equiv \pmatrix{
A&B\cr
B^{\dagger}&D
}\in\Hn{N}
\label{appham}
\ee
with the same convention given in \eqref{genham} then consider the
Feynman kernel
\be
K_k(\xi_{F},\xi_{I};T)\equiv
\langle\xi_{F};k\vert\rho_{k}(e^{-iHT})\vert\xi_{I};k\rangle\ .
\label{fkern1}
\ee
Follow a similar procedure from \eqref{characa} to \eqref{characg} to obtain
\begin{eqnarray}
&&
K_k(\xi_{F},\xi_{I};T)
\nonumber\\
& = &
{1\over[\det\{(1_{n}+\xi_{F}^{\dagger}\xi_{F})
(1_{n}+\xi_{I}^{\dagger}\xi_{I})\}]^{k/2}}
\lim\limits_{M\rightarrow\infty}
\int \prod_{i=1}^{M-1}d\mu(\xi(i);k)
\nonumber\\
&&
\times
\exp\left[{
-k\sum_{j=1}^{M-1}\tr
\log(1_{n}+\xi^{\dagger}(j)\xi(j))
+k\sum_{j=1}^{M}\tr
\log(1_{n}+\xi^{\dagger}(j)\xi(j-1))
}\right]
\nonumber\\
& &
\times
\exp\left[{
-ik\Delta t
\sum_{j=1}^{M}
\tr\{P(\xi(j),\xi(j-1))H\}
}\right] \ ,
\label{fkern2}
\end{eqnarray}
where \(\xi(0)=\xi_{I},\ \xi(M)=\xi_{F}\).
Introduce a one-parameter subgroup of \(\Un{N}\)
\be
g(t)=\exp(-iHt)=\pmatrix{
\a(t)&\b(t)\cr
\gg(t)&\d(t)
} \ , \quad t\in\bbox{R} ,
\label{1paramunit}
\ee
and write with the abbreviation \(\a(j\Dt)=\a(j)\) etc.
\be
{\bf L}(i,j) \equiv
\a(i-j)+\xi^{\dagger}(i)\gg(i-j)+\b(i-j)\xi(j)
+\xi^{\dagger}(i)\d(i-j)\xi(j)\ ,
\label{lagdef}
\ee
then discard \(O((\Delta t)^{2})\) terms to obtain
\begin{eqnarray}
&&
K_k(\xi_{F},\xi_{I};T)
\nonumber\\
& = &
{1\over[\det\{(1_{n}+\xi_{F}^{\dagger}\xi_{F})
(1_{n}+\xi_{I}^{\dagger}\xi_{I})\}]^{k/2}}
\lim\limits_{M\rightarrow\infty}
\int \prod_{i=1}^{M-1}d\mu(\xi(i);k)
\nonumber\\
&&
\times
\exp\left[{
k\sum_{j=1}^{M}\tr
\log{\bf L}(j,j-1)
-k\sum_{j=1}^{M-1}\tr
\log{\bf L}(j,j)
}\right] \ ,
\label{fkern3}
\end{eqnarray}
whose \(j\)-th integration part is
\be
\int{d\mu(\xi(j);k)\over
\{\det(1_{n}+\xi^{\dagger}(j)\xi(j))\}^{k}}
\left[
\det\left\{{{\bf L}(j+1,j){\bf L}(j,j-1)}\right\}
\right]^{k}\ .
\label{jthint1}
\ee
To carry out this integration, write
\bea
&&
\det\{{\bf L}(j+1,j){\bf L}(j,j-1)\}
\nonumber\\
&=&\det\{\a(1)+\xi^{\dagger}(j+1)\gg(1)\}
\det\left\{{1_{n}+\{\xi^{\dagger}(j+1)\ast g(1)\}\xi(j)}\right\}
\nonumber\\
&&\times
\det\left\{{1_{n}+\xi^{\dagger}(j)\{g(1)\ast \xi(j-1)\}}\right\}
\det\{\a(1)+\b(1)\xi(j-1)\}\ ,
\label{jthint2}
\eea
where
\bea
g(j)\ast \xi
& \equiv &
\{\gg(j)+\d(j)\xi\}\{\a(j)+\b(j)\xi\}^{-1}\ ,
\nonumber\\
\xi^{\dagger}\ast g(j)
& \equiv &
\{\a(j)+\xi^{\dagger}\gg(j)\}^{-1}
\{\b(j)+\xi^{\dagger}\d(j)\}\ .
\label{gact}
\eea
Utilizing the formula \eqref{kanzenseia} we find
\bea
\text{\eqref{jthint1}}
&=&
\det\{\a(1)+\xi^{\dagger}(j+1)\gg(1)\}^{k}
\nonumber\\
&&\times
\left[{\det
\left\{{1_{n}+
\{\xi^{\dagger}(j+1)\ast g(1)\}\{g(1)\ast \xi(j-1)\}}\right\}
}\right]^{k}
\nonumber\\
&&\times
\det\{\a(1)+\b(1)\xi(j-1)\}^{k}
\ ,
\label{jthint3}
\eea
which is nothing but
\be
\text{\eqref{jthint1}}
=
\left\{{
\det{\bf L}(j+1,j-1)}\right\}^{k}
\ ,
\label{jthint4}
\ee
since \(g(t)\) is an element of the one-parameter subgroup
\eqref{1paramunit}.
Hence after \(M-1\) times of this manipulation we obtain
\be
K_k(\xi_{F},\xi_{I};T)
= \left[{\det\{\a(T)+\xi^{\dagger}_{F}\gg(T)+\b(T)\xi_{I}
+\xi^{\dagger}_{F}\d(T)\xi_{I}\}\over
\det\{(1_{n}+\xi_{F}^{\dagger}\xi_{F})
(1_{n}+\xi_{I}^{\dagger}\xi_{I})\}^{1/2}}\right]^{k}\ .
\label{fkern4}
\ee

\subsection{Derivation of the Feynman kernel (method 2)}
Alternative representation can be given by the Schwinger
boson technique. First introduce an integral representation of the
inner product
between coherent states such that
\bea
\langle\xi;k\vert\eta;k\rangle
&=&
{{\cal N}(n,N-n+k)\over{\cal N}(n,k)}
\int\left({d\zeta d\bar{\zeta}\over\pi}\right)^{n^{2}}
\{\det(\zeta^{\dagger}\zeta)\}^{N-n}
\nonumber\\
&&\times
\int_{\Un{n}}{dg\over(\det g)^{k}}
\langle Z(\xi,\zeta)\vert Z(\eta,\zeta)g\rangle\ ,
\label{inpapp}
\eea
where
\be
\vert Z(\xi,\zeta)\rangle
\equiv \exp\left\{{
\tr(a^{\dagger}Z(\xi,\zeta)-Z^{\dagger}(\xi,\zeta)a)
}\right\}\vert 0\rangle \
\label{zxizeta}
\ee
is the canonical coherent state with \(Z(\xi,\zeta)\) being defined by
\bea
&&\xi\in\Mnm{N-n}{n},\ \zeta\in\Mn{n}
\nonumber\\
&\mapsto&
Z(\xi,\zeta) \equiv \pmatrix{
1_{n}\cr
\xi
}
{1\over\sqrt{1_{n}+\xi^{\dagger}\xi}}\zeta \ \in \Mnm{N}{n}\ .
\label{liftxitoz}
\eea
(The relation \eqref{inpapp} can be verified by use of the formulae
\ref{excellent1} and
\ref{excellent2}.)
Then the Feynman kernel \eqref{fkern1} is expressed as
\bea
&&
K_k(\xi_{F},\xi_{I};T)
\nonumber\\
&=&
{{\cal N}(n,N-n+k)\over{\cal N}(n,k)}
\int\left({d\zeta d\bar{\zeta}\over\pi}\right)^{n^{2}}
\{\det(\zeta^{\dagger}\zeta)\}^{N-n}
\nonumber\\
&&\times
\int_{\Un{n}}{dg\over(\det g)^{k}}
\langle Z(\xi_{F},\zeta)\vert \exp(-i\hat{H}T)
\vert Z(\xi_{I},\zeta)g\rangle\ ,
\label{skern1}
\eea
where \(\hat{H}\) has been given by \eqref{opham} for \(H\) in \eqref{appham}.
The last quantity in \eqref{skern1} is given by
\bea
&&
\langle Z(\xi_{F},\zeta)\vert \exp(-i\hat{H}T)
\vert Z(\xi_{I},\zeta)g\rangle
\nonumber\\
&=&
\exp\left\{{-\tr(\zeta^{\dagger}\zeta)}\right\}
\lim\limits_{M\rightarrow\infty}
\int\prod_{i=1}^{M-1}\left({
dz(i)d\bar{z}(i)\over\pi}\right)^{Nn}
\nonumber\\
&&
\times
\exp\left[{-\tr \left\{{
\sum_{j=1}^{M-1}Z^{\dagger}(j)Z(j)-\sum_{j=1}^{M}
Z^{\dagger}(j)g(1)Z(j-1)}\right\}
}\right]
\ ,
\label{skern2}
\eea
with
\(Z(0)=Z(\xi_{I},\zeta)g,\ Z^{\dagger}(M)=Z^{\dagger}(\xi_{F},\zeta)\),
which becomes
\be
\text{\eqref{skern2} }= \exp\left[{
-\tr\left\{{\zeta^{\dagger}\zeta
-g\zeta^{\dagger}{\bf K}(\xi_{F},\xi_{I};T)\zeta
}\right\}}\right]\ ,
\label{skern3}
\ee
where
\bea
&&{\bf K}(\xi_{F},\xi_{I};T)
\nonumber\\
&\equiv &
(1_{n}+\xi_{F}^{\dagger}\xi_{F})^{-1/2}
{\bf L}(M,0)
(1_{n}+\xi_{I}^{\dagger}\xi_{I})^{-1/2}
\nonumber\\
&=&
(1_{n}+\xi_{F}^{\dagger}\xi_{F})^{-1/2}
\{\a(T)+\xi^{\dagger}_{F}\gg(T)+\b(T)\xi_{I}
+\xi^{\dagger}_{F}\d(T)\xi_{I}\}
(1_{n}+\xi_{I}^{\dagger}\xi_{I})^{-1/2}\ .
\nonumber\\
\label{matkern}
\eea
Substituting \eqref{skern3} into \eqref{skern1} and carrying out
integrations with respect to \(g\) and \(\zeta\), we find
\bea
K_k(\xi_{F},\xi_{I};T) &= &\{\det{\bf K}(\xi_{F},\xi_{I};T)\}^{k}
\nonumber \\
&= & \left[{\det\{\a(T)+\xi^{\dagger}_{F}\gg(T)+\b(T)\xi_{I}
+\xi^{\dagger}_{F}\d(T)\xi_{I}\}\over
\det\{(1_{n}+\xi_{F}^{\dagger}\xi_{F})
(1_{n}+\xi_{I}^{\dagger}\xi_{I})\}^{1/2}}\right]^{k}
\label{skern5}
\eea
which of course matches to the result obtained in the previous section,
however,
this rewrite can again provide an interpretation of the
WKB exactness.

\subsection{The WKB approximation}
{}From the path integral expression
\eqref{fkern3}, the action is given by
\be
S= \sum_{j=1}^{M}\tr
\log{\bf L}(j,j-1)
- \sum_{j=1}^{M-1}\tr
\log{\bf L}(j,j)\ ,
\label{pathact}
\ee
and therefore equations of motion read
\bea
\{\gg(1)+\d(1)\xi(j-1)\}
{\bf L}^{-1}(j,j-1)
&=&
\xi(j){\bf L}^{-1}(j,j)\ ,
\label{eqomappa}\\
{\bf L}^{-1}(j+1,j)\{\b(1)+\xi^{\dagger}(j+1)\d(1)\}
&=&
{\bf L}^{-1}(j,j)\xi^{\dagger}(j)\ ,
\label{eqomappb}
\eea
for \(1\le j\le M-1\).
These can be solved locally by
\bea
\xi(j)&=&\{\gg(1)+\d(1)\xi(j-1)\}
\{\a(1)+\b(1)\xi(j-1)\}^{-1}\ ,
\label{clsl1a}\\
\xi^{\dagger}(j)&=&\{\a(1)+\xi^{\dagger}(j+1)\gg(1)\}^{-1}
\{\b(1)+\xi^{\dagger}(j+1)\d(1)\}\ ,
\label{clsl1b}
\eea
so that the classical solution is, after taking account of the boundary
condition
\(\xi(0) = \xi_{I}; \xi(M) = \xi_{F} \), found such that
\bea
\xi_{c}(j)
&=&\{\gg(j)+\d(j)\xi_{I}\}
\{\a(j)+\b(j)\xi_{I}\}^{-1}\ ,
\label{clsl2a}\\
\xi_{c}^{\dagger}(j)&=&\{\a(M-j)+\xi^{\dagger}_{F}\gg(M-j)\}^{-1}
\{\b(M-j)+\xi^{\dagger}_{F}\d(M-j)\}\ .
\label{clsl2b}
\eea

Here two comments are in order: the first is on the fact that
\(\xi_{c}(j)\) and \(\xi_{c}^{\dagger}(j)\)
are not Hermitian conjugated each other.
This is not so surprising, although it is often emphasized.
A similar situation can be met even in a simple Gaussian integration:
\be
\int{dzd\bar{z}\over\pi}\exp(-\bar{z}z+\bar{a}z+\bar{z}b)\ ,
\label{examp1}
\ee
for \(a,b\in\bbox{C}\). Complete the square to find
\be
\bar{z}z-\bar{a}z-\bar{z}b=
(\bar{z}-\bar{a})(z-b)-\bar{a}b\ ,
\label{heihou}
\ee
then shift the variable \(z(\bar{z})\) by the amount of \(b(\bar{a})\) to
obtain
\be
\int{dzd\bar{z}\over\pi}\exp(-\bar{z}z+\bar{a}b)=e^{\bar{a}b}\ .
\label{examp2}
\ee
The shift of \(z\) and \(\bar{z}\) is asymmetric but there is no problem.
The point is that they
should be regarded independent. In our case  we have two  independent
sources \(\xi_{I}\) and \(\xi^{\dagger}_{F}\) which are not always
Hermitian conjugated each
other.

The second remark is on the fact that there is no
overspecification problem which is also often stressed by authors in the
frame work of
continuum  version of path integral; since it seems two
boundary conditions in
the first order differential equation. But under the discrete time
formalism there is no room for
appearance of \(\xi^{\dagger}_{I}\) and \(\xi_{F}\) as to the boundary
conditions.

Now turn back to the main subject: we must evaluate the action
\eqref{pathact} and its Hessian at classical solutions
\eqref{clsl2a} and \eqref{clsl2b}.
To this end note the following relation:
\be
{\bf L}(j,j)_{c}\{\a(j)+\b(j)\xi_{I}\}
={\bf L}(j,j-1)_{c}\{\a(j-1)+\b(j-1)\xi_{I}\}\ ,
\label{benri1}
\ee
where the subscript \(c\) designates quantities of the classical solutions.
By use of this relation, we can easily evaluate the classical action:
\bea
S_{c}
& \equiv &
 \sum_{j=1}^{M}\tr\log{\bf L}(j,j-1)_{c}
- \sum_{j=1}^{M-1}\tr\log{\bf L}(j,j)_{c}
\nonumber\\
& = &
 \tr\log
\left[{{\bf L}(M,M-1)_{c}\{\a(M-1)+\b(M-1)\xi_{I}\}}\right]
\nonumber\\
& = &
 \tr\log
{\bf L}(M,0)\ ,
\label{clasact}
\eea
with \({\bf L}(M,0)\in\Mn{n}\) being given by \eqref{lagdef}: explicitly
\(
{\bf L}(M,0)
=\a(T)+\xi^{\dagger}_{F}\gg(T)+\b(T)\xi_{I}
+\xi^{\dagger}_{F}\d(T)\xi_{I}
\).
Let us introduce another one-parameter subgroup by
\be
\tilde{g}(t)\equiv
\pmatrix{
\d^{\dagger}(t)&-\b^{\dagger}(t)\cr
-\gg^{\dagger}(t)&\a^{\dagger}(t)
}\ ,
\label{anonpsbg}
\ee
to define
\be
\tilde{\bf L}(i,j)
\equiv
\d^{\dagger}(i-j)-\xi(j)\gg^{\dagger}(i-j)
-\b^{\dagger}(i-j)\xi^{\dagger}(i)
+\xi(j)\a^{\dagger}(i-j)\xi^{\dagger}(i)\ .
\label{ltilde}
\ee
Then the Hessian of the action is found as
\be
{\partial^{2}S\over\partial\xi_{ia}(l)\partial\bar{\xi}_{jb}(m)}
=-\d_{l,m}
\tilde{\bf L}^{-1}(m,m)_{ji}{\bf L}^{-1}(m,m)_{ab}
+\d_{l,m-1}\tilde{\bf L}^{-1}(m,m-1)_{ji}
{\bf L}^{-1}(m,m-1)_{ab}\ .
\label{2ndderiv}
\ee
Shifting the variables
\be
\xi(j)= \xi_{c}(j)+z(j),\ \xidag(j)=\xidag_{c}(j)+z^{\dagger}(j) \ ,
\label{expansion}
\ee
we approximate the action up to the bilinear terms of \(z\) and
\(z^\dagger\):
\bea
S
& \sim &
S_{c}
- \sum_{j=1}^{M-1}\tr
\left\{{ z^{\dagger}(j)\tilde{\bf L}^{-1}(j,j)_{c}z(j)
{\bf L}^{-1}(j,j)_{c}}\right\}
\nonumber\\
& &
+ \sum_{j=2}^{M-1}\tr
\left\{{ z^{\dagger}(j)\tilde{\bf L}^{-1}(j,j-1)_{c}z(j-1)
{\bf L}^{-1}(j,j-1)_{c}}\right\}\ ,
\label{bilinear}
\eea

The measure is also approximated as
\be
\left.{\vphantom{\prod_{i=1}^{M}}
d\mu(\xi(j);k)}\right\vert_{c}
\stackrel{k \rightarrow \infty}{\sim}
{k^{n(N-n)}\over{\{\det{\bf L}(j,j)_{c}\}^{N}}}
\left({d z(j)d\bar{z}(j)\over\pi}\right)^{n(N-n)}
\left\{1+O(k^{-1})\right\} \ ,
\label{measurecl}
\ee
where use has been made of the definition of the measure \eqref{invmeas}
with \eqref{largek}.
Therefore in view of \eqref{bilinear} and \eqref{measurecl} the Gaussian
integration of \(\bar{z}(j), z(j)\) gives
\be
\det\left\{{\tilde{\bf L}^{-1}(j,j)_{c}\otimes
({\bf L}^{-1}(j,j)_{c})^{\cal T}}\right\}^{-1}
=
\{\det{\bf L}(j,j)_{c}\}^{N}
\label{voldet}
\ee
which together with \(\left( \pi /  k \right)^{n(N-n)}\) cancels the
prefactor in \eqref{measurecl}, yielding the leading contribution to the WKB
approximation:
\bea
K_k(\xi_{F},\xi_{I};T)&\stackrel{k \rightarrow \infty}{\sim}&
{1\over
[\det\{(1_{n}+\xi_{F}^{\dagger}\xi_{F})
(1_{n}+\xi_{I}^{\dagger}\xi_{I})\}]^{k/2}}
\exp\{k\tr\log{\bf L}(M,0)\}
\nonumber\\
&=&
\left[{\det\{\a(T)+\xi^{\dagger}_{F}\gg(T)+\b(T)\xi_{I}
+\xi^{\dagger}_{F}\d(T)\xi_{I}\}\over
\det\{(1_{n}+\xi_{F}^{\dagger}\xi_{F})
(1_{n}+\xi_{I}^{\dagger}\xi_{I})\}^{1/2}}\right]^{k}\ .
\label{WKBkernel}
\eea
The result is equal to the full form of the Feynman kernel \eqref{fkern4},
\eqref{skern5}. Thus {\em the WKB approximation is again exact for the Feynman
kernel}, which can also be clarified in terms of \eqref{skern1} as well as
\eqref{skern2}; since the integrand is Gaussian.

%%%%%% Biblography %%%%%

\end{document}